\documentclass[useAMS,usenatbib]{mn2e}
\usepackage{graphicx}
\usepackage{natbib}
\usepackage{lscape}
\usepackage{longtable}
\usepackage{subfig}
\usepackage{amssymb,amsmath}
\usepackage[T1]{fontenc}
\usepackage{hyperref}
\usepackage{amsmath}
\usepackage{breakurl}
\usepackage{times}
\usepackage{morefloats}
\usepackage{bm}
\usepackage{flafter}

\newcommand{\wise}{\textit{WISE}}

\newcommand\ion[2]{#1$\;${\small\rmfamily\@Roman{#2}}\relax}%

\def\lsim{\lower0.3em\hbox{$\,\buildrel <\over\sim\,$}}
\def\gsim{\lower0.3em\hbox{$\,\buildrel >\over\sim\,$}}

\title[NTO quasar halo masses]{The Impact of the Dusty Torus on Obscured Quasar Halo Mass Measurements}
\author[DiPompeo et al.]{M.A. DiPompeo$^{1,2}$, J.C. Runnoe$^3$, R.C. Hickox$^1$, A.D. Myers$^2$, J.E. Geach$^4$ \\
$^1$ Department of Physics and Astronomy, Dartmouth College, 6127 Wilder Laboratory, Hanover, NH 03755, USA  \\
$^2$ Department of Physics and Astronomy 3905, University of Wyoming, 1000 E. University, Laramie, WY 82071, USA \\
$^3$ Department of Astronomy \& Astrophysics, and Institute for Gravitation and the Cosmos, The Pennsylvania State University, 525 Davey Lab, University Park, PA 16802, USA \\
$^4$ Centre for Astrophysics Research, Science \& Technology Research Institute, University of Hertfordshire, Hatfield, AL10 9AB, UK}

\begin{document}
\date{Accepted 22 April 2016; Submitted 1 February 2016}

\pagerange{\pageref{firstpage}--\pageref{lastpage}} \pubyear{2016}

\maketitle

\label{firstpage}

\begin{abstract}
Recent studies have found that obscured quasars cluster more strongly and are thus hosted by dark matter haloes of larger mass than their unobscured counterparts.  These results pose a challenge for the simplest unification models, in which obscured objects are intrinsically the same as unobscured sources but seen through a dusty line of sight.  There is general consensus that a structure like a ``dusty torus'' exists, meaning that this intrinsic similarity is likely the case for at least some subset of obscured quasars.  However, the larger host halo masses of obscured quasars implies that there is a second obscured population that has an even higher clustering amplitude and typical halo mass.  Here, we use simple assumptions about the host halo mass distributions of quasars, along with analytical methods and cosmological $N$-body simulations to isolate the signal from this population.  We provide values for the bias and halo mass as a function of the fraction of the ``non-torus obscured'' population.  Adopting a reasonable value for this fraction of $\sim$25\% implies a non-torus obscured quasar bias that is much higher than the observed obscured quasar bias, because a large fraction of the obscured population shares the same clustering strength as the unobscured objects.  For this non-torus obscured population, we derive a bias of $\sim$3, and typical halo masses of $\sim3\times10^{13}$ M$_{\odot}/h$ at $z=1$.  These massive haloes are likely the descendants of high-mass unobscured quasars at high redshift, and will evolve into members of galaxy groups at $z=0$.\end{abstract}


\section{INTRODUCTION}
Recent analyses of large samples of obscured quasars, those with very red infrared (IR) and optical-IR colours, have uncovered an intriguing trend that indicates a departure from the simplest unification-by-orientation models \citep[e.g.][and references therein]{1993ARA&A..31..473A, 2015ARA&A..53..365N}: they seem to reside, on average, in dark matter haloes of higher mass than unobscured quasars selected in a similar way.  In the case of an axis-symmetric obscurer (the ``dusty torus'') and random orientations, one expects that the global average properties of obscured and unobscured quasars are the same.  A higher halo mass in obscured sources suggests additional factors at work.

The study of halo masses through correlation analyses requires large statistical samples, and in the era of large surveys has been a topic of rigorous study for optically bright unobscured quasars \citep[e.g.][]{2004MNRAS.355.1010P, 2005MNRAS.356..415C, 2007ApJ...654..115C, 2007ApJ...658...85M, 2008MNRAS.383..565D, 2009MNRAS.397.1862P, 2009ApJ...697.1634R, 2010ApJ...713..558K, 2012MNRAS.424..933W, 2013ApJ...778...98S, 2015MNRAS.453.2779E}.  The measurement is typically made using either the quasar autocorrelation function or a cross-correlation with galaxies.  By comparing this signal with that expected from dark matter in a given cosmological model, the excess amplitude of the auto/cross-correlation (which provides the quasar bias, $b$) can be inferred and converted into a typical halo mass for the sample.

Selecting large samples of obscured quasars efficiently is currently only possible in the mid-IR \citep{2004ApJS..154..166L, Stern:2005p2563, 2012ApJ...753...30S, 2013ApJ...772...26A, 2013MNRAS.434..941M}.  Only in the last several years, first with samples selected with \textit{Spitzer} \citep[][]{2004ApJS..154....1W} and more recently with the \textit{Wide-field Infrared Survey Explorer} \citep[\wise;][]{2010AJ....140.1868W}, have clustering analyses of obscured quasars been performed.  While the results have varied as samples are refined, the general consensus is that the average bias, and thus halo mass, of obscured quasars is higher \citep[][though see also \citealt{2015arXiv150406284M}]{2011ApJ...731..117H, 2014MNRAS.442.3443D, 2016MNRAS.456..924D}.  Recent follow-up using gravitational lensing maps of the cosmic microwave background \citep[CMB;][]{2012ApJ...756..142V, 2014A&A...571A..17P, PlanckCollaboration:2015tp} to directly probe the masses of quasar hosts have generally confirmed clustering measurements \citep[e.g.][]{2012PhRvD..86h3006S}, including the higher halo masses of obscured quasars \citep[][though see also \citealt{2013ApJ...776L..41G}]{2015MNRAS.446.3492D, 2016MNRAS.456..924D}.

The dusty torus model successfully explains many observed properties of low-$z$, low-$L$ active galactic nuclei \citep[AGN, e.g.][]{1993ARA&A..31..473A, Urry:1995p507}, with additional factors at play, such as accretion rate and luminosity \citep[e.g.][]{2000ApJ...540L..73M, 2007ApJ...657..159B, 2012ApJ...748..130M}.  However, the ability of the dusty torus model to explain the full AGN population is less certain for higher redshift quasars.  The presence of hot dust in the nuclear regions of quasars is established, as it is what allows their selection from the signature red power-law of accretion disk heated dust in the IR.  This dust --- including its geometry, column density, chemical make up, and origin --- is an area of vigorous study for theorists and observers alike (see sections 5.1.1 and 5.1.2).  Regardless of the details, it is very likely that all IR-selected unobscured quasars would be seen as obscured along the appropriate line of sight due to nuclear dust, even the so-called hot-dust-poor quasars \citep{2011ApJ...733..108H}.  We will refer to such objects as torus-obscured quasars throughout. 

There are also other ways to obscure the quasar activity in the nuclei of galaxies.  For example, large-scale galactic dust has been identified as a significant source of obscuration in some sources, such as those in galaxies seen edge-on or with powerful starbursts \citep[e.g.][]{2012ApJ...755....5G, 2015ApJ...802...50C}.  This kind of obscuration is predicted in some models of black hole and galaxy coevolution, some of which invoke mergers as a driver for the most powerful quasar activity \citep[e.g.][]{1988ApJ...325...74S, 2008ApJS..175..356H, 2009MNRAS.394.1109C, 2010MNRAS.405L...1B, 2010MNRAS.407.1529H}.  These scenarios can also impact nuclear dust, stirring it up into different configurations and implying evolution of obscuration on large and small scales.  

In the case of large-scale dust, which falls well outside the size scales of the quasar narrow-line region, there may be differences in the optical spectral features compared to torus-obscured sources where the obscuration occurs interior to the narrow-line region.  Spectroscopic follow-up of quasars selected based on their mid-IR colours shows that while a large fraction have AGN-dominated narrow emission lines, there is a substantial fraction ($\sim$$10-20$\%) that do not \citep{2013ApJS..208...24L, 2014ApJ...795..124H}.  Despite this, many are X-ray sources indicative of nuclear activity, which may indicate that the narrow lines are being obscured as well as the continuum.

All of these results indicate that in any obscured quasar sample, there is potentially a mix of objects that are intrinsically like the unobscured objects, and thus obscured by only a torus, along with sources that are obscured by some other factor (galaxy-wide dust, nuclear dust in some other geometry or evolutionary state).  This is true for the bias and halo mass studies above, and implies that the measured obscured quasar bias is in fact a lower limit on the bias of objects that are intrinsically different from unobscured sources, which we will refer to as non-torus-obscured (NTO) quasars.  These NTO quasars likely also contain nuclear dust, and may potentially be obscured by a torus in addition to other sources of obscuration for certain lines of sight --- nevertheless, they represent a distinct population that is unique because of the obscuring material not associated with the torus.  Assuming that the torus-obscured sources cluster like the unobscured quasars, if we knew the fraction of NTO objects we could separate the NTO bias, and thus halo masses, shedding additional light on the properties of this important quasar class.  That is the goal of this work.

In section 2 we provide a summary of the observational constraints used in our analysis.  In section 3 we provide a purely analytical approach to separating the NTO bias, and in section 4 we explore these analytical predictions using samples drawn from cosmological simulations.  In section 5 we turn to observations and models that may constrain the NTO fraction, and use these to speculate on the nature of the NTO population.

\section{OBSERVATIONAL CONSTRAINTS}
Our modeling is constrained by the bias and halo mass measurements of \citet{2016MNRAS.456..924D}, hereafter D16, which were made by cross-correlating CMB lensing maps from \textit{Planck} \citep{PlanckCollaboration:2015tp} and maps of the relative density of quasars selected with \wise.  Though we will use simulated measurements of the angular autocorrelation function in section 4, a measurement also made by D16, the bias measurements from the CMB lensing cross-correlations are likely more reliable.  In any case, the two methods agree quite well, so adopting one or the other will not strongly affect our results here.  We refer the reader to D16 for complete details, but highlight a few aspects of their data here.

The quasars in D16 are selected based on \wise\ IR colours, using the simple colour cut of $W1-W2 > 0.8$ and a magnitude limit of $W2 < 15.05$ \citep{2012ApJ...753...30S}\footnote{$W1$ and $W2$ refer to the \wise\ filters centered on 3.5 and 4.6 $\mu$m, respectively.  The native \wise\ system is Vega magnitudes.}.  These are carefully masked to limit artifacts and other sources of contamination (i.e.\ scattered Moonlight, regions of high Galactic extinction, bright stars, etc.).  Only sources that meet these criteria in both versions of the \wise\ catalogues (ALLSKY and ALLWISE) are included in the final sample, as a conservative approach.  The samples are matched to imaging from the Sloan Digital Sky Survey \citep{2000AJ....120.1579Y}, and optical to IR colours are used to separate obscured and unobscured sources \citep[with a split at $r-W2 = 6$, in Vega magnitudes,][]{2007ApJ...671.1365H}.  Objects without optical counterparts are placed in the obscured sample.

The final sample is 141,875 quasars and covers an area of 2994 deg$^2$.  The unobscured to obscured ratio is approximately 60/40, implying source densities of $\sim$20 deg$^{-2}$ (obscured) and $\sim$30 deg$^{-2}$ (unobscured).  The mean redshifts of each sample are 0.96 and 1.05 for obscured and unobscured quasars, respectively, each with a standard deviation of $\sim$0.5.  Using a model based on the cosmology of \citet{2011ApJS..192...18K}, D16 measured biases of $b_{\textrm{obsc}}=2.06\pm0.22$ and $b_{\textrm{unob}} = 1.72\pm0.18$.

When utilizing simulated halo catalogues in Section 4, we are bound to the choices of cosmological parameters of the simulations.  Our simulation of choice is \textsc{MultiDark}, which we will introduce and describe fully in section 4.1.  Here we update the bias measurements of D16 by refitting their data using a model consistent with the \textsc{MultiDark} cosmology (see section 3.3 of D16 and the supplied code library); $H_0 =70$ km/s/Mpc, $\Omega_m = 0.27$, $\Omega_{\Lambda} = 0.73$, $\Omega_b = 0.0469$, along with $\sigma_8 = 0.82$ and a matter power spectrum spectral index of $n=0.95$ (D16 used $n=0.96$).  These changes are small, and within the errors associated with these parameters, but we utilize them nonetheless.  These updates change the measured biases minimally to $b_{\textrm{obsc}} = 2.02\pm0.2$ and $b_{\textrm{unob}}=1.68\pm0.17$.

\section{AN ANALYTICAL APPROACH}
\subsection{The halo mass distributions}
Given the mass distribution of dark matter haloes ($dN/dM$), combined with a model for the linear bias as a function of halo mass $b(M)$, the mass-averaged bias can be determined:
\begin{equation}
\label{eq:beff}
b = \frac{\int b(M) \frac{dN}{dM} dM}{\int \frac{dN}{dM}dM}.
\end{equation}
There are several parameterizations of $b(M)$, from both dark matter collapse models as well as $N$-body simulations \citep[e.g.][]{2001MNRAS.323....1S, 2005ApJ...631...41T, 2010ApJ...724..878T}.  We adopt the parameterization of \citet[][see also section 3.5 of D16]{2010ApJ...724..878T}.  The bias is an evolving function of redshift, as gravity produces more structure as the Universe ages.  In the absence of individual source redshifts, it is common to analyze the bias and halo masses as ``effective'' values taken at the mean redshift of the sample, as in \citet[][]{2014MNRAS.442.3443D, 2015MNRAS.446.3492D} and D16.  Here, we will be working at $z=1$, the approximate mean of the observed obscured and unobscured samples.  Since the redshift distributions of the full obscured and unobscured samples are similar (D16), and by definition the torus-obscured objects should have the same distribution as the unobscured sources, it is reasonable to assume the non-torus-obscured objects have a similar distribution as well.

Equation~\ref{eq:beff} can be used to provide a non-parametric analytical description of the non-torus obscured bias ($b_{\textrm{NTO}}$) as a function of the NTO fraction of the obscured population ($f_{\textrm{NTO}}=N_{\textrm{NTO}}/N_{\textrm{obsc}}$).  We first split the obscured mass distributions into torus obscured (TO) and NTO components (note that for brevity we will assume the mass distributions are already normalized):
\begin{equation}
\begin{split}
b_{\textrm{obsc}} = \int b(M) \left( \frac{dN_{\textrm{NTO}}}{dM} + \frac{dN_{\textrm{TO}}}{dM} \right) dM  \\
=b_{\textrm{NTO}}f_{\textrm{NTO}} + b_{\textrm{TO}}(1-f_{\textrm{NTO}}).
\end{split}
\end{equation}
In our model, $b_{\textrm{TO}}$ is equivalent to the observed unobscured bias, and so this can be rearranged as
\begin{equation}
b_{\textrm{NTO}} = \frac{b_{\textrm{obsc}} - b_{\textrm{unob}}}{f_{\textrm{NTO}}} + b_{\textrm{unob}}.
\label{eq:bnto}
\end{equation}
In order to give a more physical basis to our model, and to incorporate what is already known about the halo mass distributions of unobscured quasars, in what follows we will work with a parametric form of $dN/dM$ and Equation~\ref{eq:beff}.  However, Equation~\ref{eq:bnto} highlights the fact that our results are not dependent on the details of the form of mass distributions.

Starting from the mean occupation function of quasars, which describes the probability of finding $N$ quasars in a halo of a given mass, and combining this with the overall halo mass function, the mass distribution of quasar hosts can be inferred. This type of parametrization has been studied in detail for galaxies \citep{2002ApJ...575..587B, 2003ApJ...593....1B, 2005ApJ...633..791Z, 2008ApJ...682..937B} and more recently for low luminosity AGN and unobscured quasars \citep{2011ApJ...726...83M, 2012MNRAS.424..933W, 2012MNRAS.419.2657C, 2012ApJ...755...30R, 2013ApJ...779..147C, 2013ApJ...774..143R}.  These studies generally find that the quasar halo mass distribution is approximately log-normal \citep[e.g.][]{2012ApJ...755...30R}, and we adopt this parameterization here.  

While there is some suggestion that the AGN halo mass distribution evolves with redshift at low-luminosity, there is no concrete evidence of such evolution for quasars \citep{2012MNRAS.419.2657C}.  Most measurements of the typical halo masses of unobscured quasars (ignoring the underlying \textit{shape} of the mass distribution) find that it is roughly constant with redshift (see references in the introduction), which may suggest a stable distribution shape with cosmic time.  As the halo mass distribution has not been studied directly for obscured quasars, we will assume that it is also log-normal as there is no evidence to the contrary.  A log-normal parameterization has two free parameters --- the mean $\mu$ and standard deviation $\sigma$ in log-space.  

\begin{figure}
\centering
\vspace{0.3cm}
\hspace{0cm}
   \includegraphics[width=7.5cm]{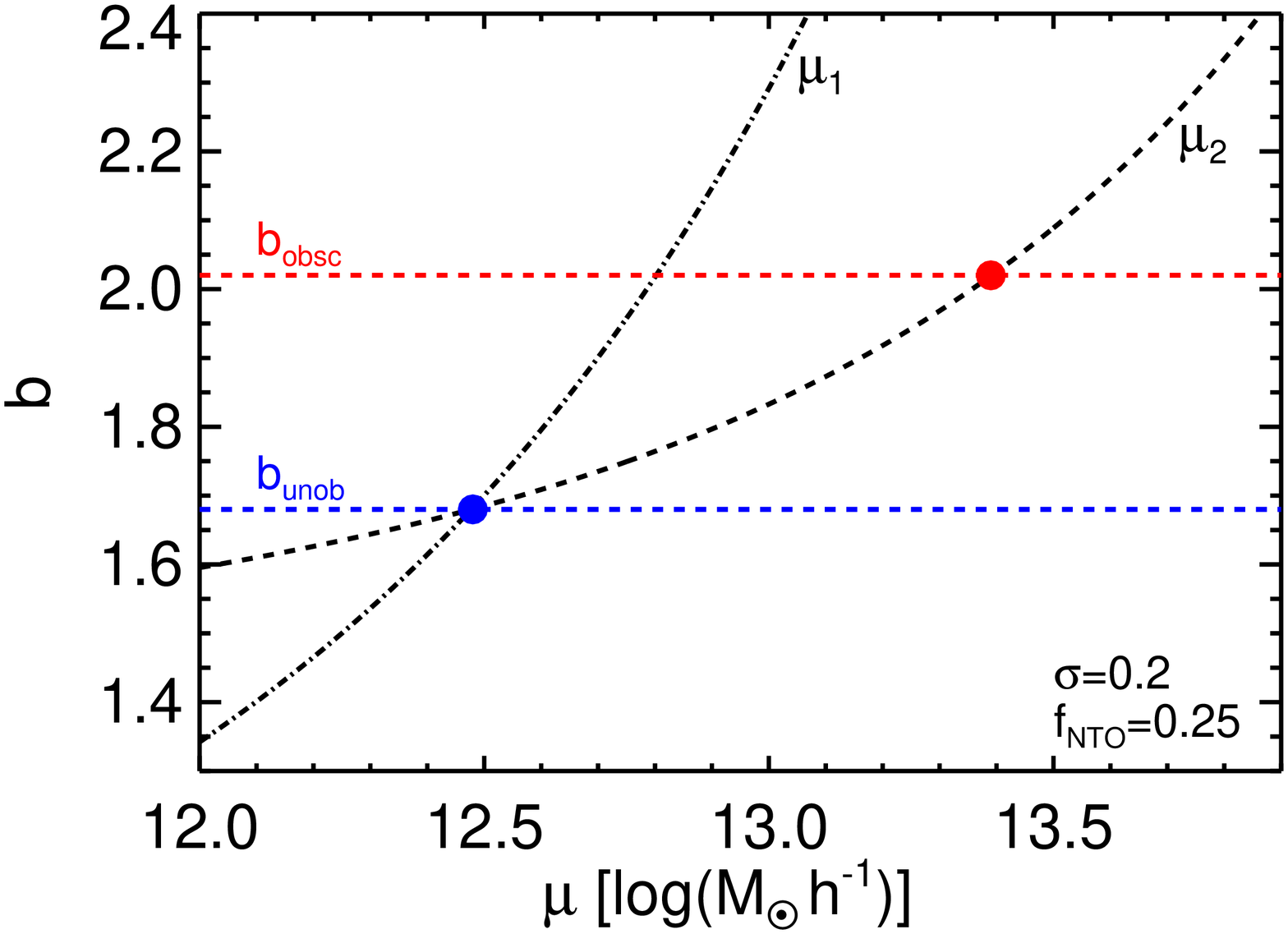}
    \vspace{0cm}
  \caption{The effective bias as a function of the mean of the halo mass distributions (using Equation~\ref{eq:beff} to integrate the log-normal distributions).  The dash-dotted line shows how shifting $\mu_1$ affects the unobscured bias, as the peak of the single distribution shifts.  For a given width ($\sigma=0.2$ here), the blue point at the intersection with the blue line marking the observed unobscured bias sets the adopted value of $\mu_1$.   The dashed line shows the effect of shifting the mean of the non-torus obscured part of the obscured distribution, while holding $\mu_1$ fixed (for $f_{\textrm{NTO}}=0.25$ here).  The red point at the intersection with the red line, marking the observed obscured bias, sets the value of $\mu_2$. We repeat this process for different values of $f_{\textrm{NTO}}$ to determine the adopted mass distributions.\label{fig:mu_b}}
\end{figure}

Since the unobscured and torus-obscured objects are assumed to be intrinsically identical, with the only difference being viewing angle, the mean and standard deviation of their halo mass distributions will always be the same in our model.  We will refer to the mean halo mass of these samples as $\mu_1$ throughout.  Using Equation 1 with a given $\sigma$, $\mu_1$ is determined by shifting the distribution until the bias matches that of the observed unobscured sample (Figure~\ref{fig:mu_b}).  

\begin{figure}
\centering
\vspace{0.3cm}
\hspace{0cm}
   \includegraphics[width=7.5cm]{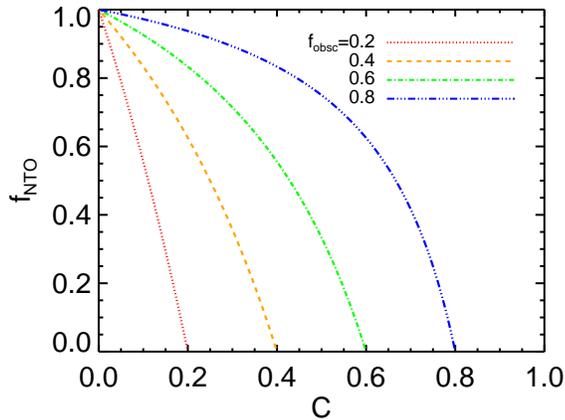}
    \vspace{0cm}
  \caption{The relationship between the torus covering factor $C$ and the non-torus obscured fraction $f_{\textrm{NTO}}$, for different values of the observed total obscured fraction $f_{\textrm{obsc}}$ (see Equation~\ref{eq:cf}).  This allows our results as a function of $f_{\textrm{NTO}}$ to be translated to other samples with different $f_{\textrm{obsc}}$ and assumed $C$.  Our sample has $f_{\textrm{obsc}}=0.4$.\label{fig:cf}}
\vspace{0.2cm}
\end{figure}

Because the halo mass distribution of the non-torus obscured sample is unknown, we also assume it is log-normal, with mean $\mu_2$ and a standard deviation tied to that of the other samples.  This means that the complete obscured halo mass distribution is a linear combination of two log-normal components with their ratios set by $f_{\textrm{NTO}}$, a free parameter.  This fraction can be related to the torus covering factor $C$ for a given value of the overall obscured fraction $f_{\textrm{obsc}}$ (which includes torus-obscured and NTO obscured objects, as well as those potentially obscured by both) by:
\begin{equation}
C = f_{\textrm{obsc}} \frac{1-f_{\textrm{NTO}}}{1-f_{\textrm{obsc}}f_{\textrm{NTO}}}.
\label{eq:cf}
\end{equation}
Note that this is the covering factor of the dust associated just with the torus, and is not the overall covering fraction that would be derived from obscured to unobscured ratios, which may include dust in a range of locations \citep[e.g.][]{2008ApJ...675..960P, 2012ApJ...755....5G}.  This relationship accounts for the fact that some lines of sight may pass through both a non-torus obscurer and a torus (i.e. they can appear to overlap), and non-torus obscured sources are just as likely as the unobscured-like population to be seen through a torus in addition to the extra obscurer.  It further assumes that there is no alignment between the angular momentum axes of quasars and their hosts \citep[e.g.][]{2000ApJ...537..152K, 2009ApJ...699..281Z}.  This covering factor is shown for a range of $f_{\textrm{obsc}}$ in Figure~\ref{fig:cf} to facilitate combining our results with other samples or models of dust geometries.  We will generally use our observed $f_{\textrm{obsc}}=0.4$ throughout, based on the observed obscured and unobscured number densities (D16).  For a given $f_{\textrm{obsc}}$ and torus model, C or $f_{\textrm{NTO}}$ can be converted into other torus properties, such as the half opening angle, which we will explore in section 5.1.  

\begin{figure}
\centering
\vspace{0.3cm}
\hspace{0cm}
   \includegraphics[width=8.5cm]{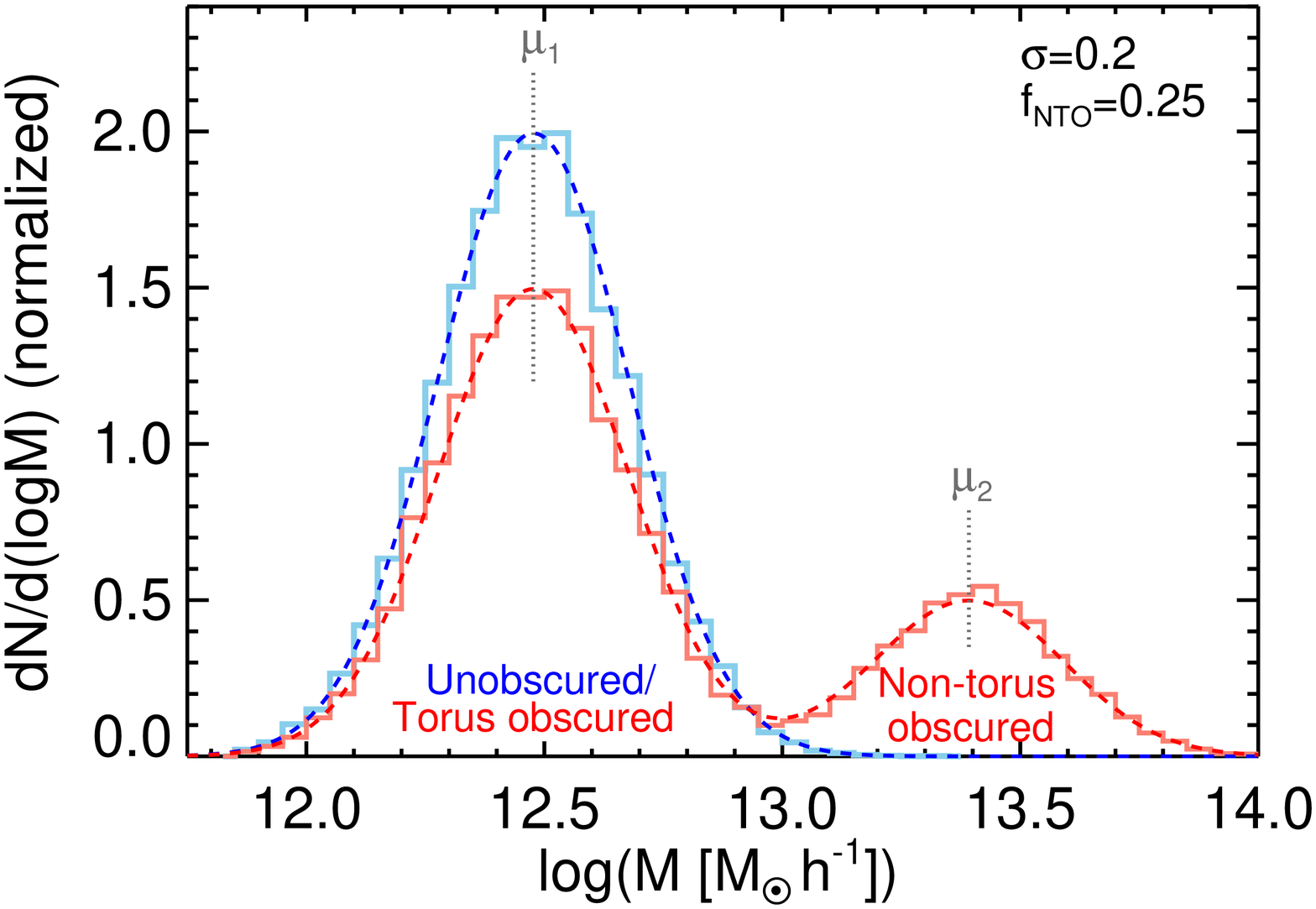}
    \vspace{0cm}
  \caption{Example mass distributions ($dN/dM$) for the unobscured (blue) and obscured (red) populations. The unobscured sources have a log-normal distribution with mean $\mu_1$ and standard deviation $\sigma=0.2$.  The obscured objects have a torus-obscured subset with the same properties as the unobscured objects, and a non-torus obscured subset with a log-normal mass distribution with mean $\mu_2$ and the same $\sigma$.  The ratio of the areas under the obscured log-normal components in this example is $f_{\textrm{NTO}}=0.25$.  We explore the role of changing this ratio, which is constrained by an observed obscured/unobscured ratio and assumed torus covering factor or torus model in section 5.1.  The dashed lines are the theoretical distributions for our analytical analysis (section 3.1), and the solid histograms show random samplings of haloes from the \textsc{MultiDark} simulations (section 4.2).\label{fig:dndm}}
\end{figure}

Once $\mu_1$ is determined, for a given $f_{\textrm{NTO}}$ we shift $\mu_2$ until the integrated bias matches the observed obscured bias (Figure~\ref{fig:mu_b}).  Figure~\ref{fig:dndm} shows an example halo mass distribution for the obscured and unobscured samples, for $\sigma=0.2$ and $f_{\textrm{NTO}}=0.25$.

\subsection{Results}
Once $\mu_2$ is determined, we can isolate the bias of the non-torus obscured sample by integrating only this second component of the obscured halo mass distribution.  We refer to this as $b_{\textrm{NTO}}$.  In practice, if we measured this bias for a sample with mean $z=1$, we would convert it into a ``typical'' halo mass for the sample, with no assumptions about the underlying distribution.  We label this mass $M_{\textrm{NTO}}$.  This value is not necessarily the same as $\mu_2$, the peak mass of the distribution, but these values approach each other in the limit of $\sigma=0$.

\begin{figure}
\centering
\vspace{0.3cm}
\hspace{0cm}
   \includegraphics[width=6.5cm]{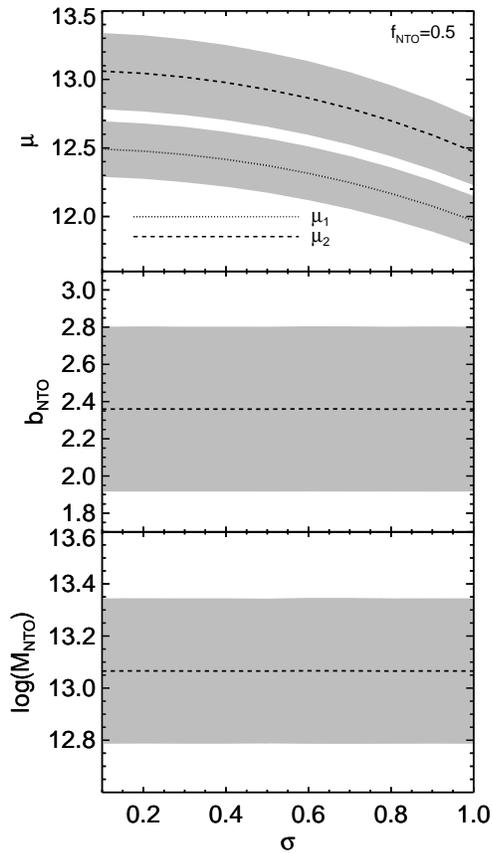}
    \vspace{0cm}
  \caption{The effect of the width of the mass distributions $\sigma$ (for fixed $f_{\textrm{NTO}}=0.5$) on the adopted $\mu$  values and resulting non-torus obscured bias and halo mass.  The gray bands show the 68\% confidence intervals based on Monte-Carlo resampling the observed biases within their errors.  \emph{Top:} The values of $\mu_1$ and $\mu_2$ required to reproduce the observed unobscured and obscured clustering amplitudes. \emph{Middle:} The inferred bias of the non-torus obscured samples, and \emph{bottom:} the ``typical'' halo mass of non-torus obscured quasars, inferred from the bias as would be done for observations where the full halo mass distribution is unknown. As expected from Equation~\ref{eq:bnto}, the bottom two relationships are flat with respect to $\sigma$.\label{fig:sig_trends}}
\vspace{0.2cm}
\end{figure}

Equation~\ref{fig:sig_trends} shows that the specific shape of the halo mass distributions will not impact the results, and this implies that the exact value of $\sigma$ is not important.  We verify that this is the case, as shown in Figure~\ref{fig:sig_trends}.  The top panel shows that indeed the mean halo masses $\mu_1$ and $\mu_2$ do depend on $\sigma$ (which reflects the shape of $b(M)$, and the fact that a wider distribution includes more massive haloes with disproportionately large bias), but the bottom two panels highlight that the effective bias and halo mass of the NTO population does not.  The grey regions indicate the 68\% confidence intervals calculated by randomly sampling the observed obscured and unobscured biases from distributions consistent with their measured errors.  Because of the lack of dependence on $\sigma$, we fix this width for all samples at 0.2 for the remainder of this analysis.  This value is not chosen for any physical reason, though it does make sampling haloes from simulations (see section 4) somewhat easier as we do not have to obtain simulated haloes for as wide of a mass range.

\begin{figure}
\centering
\vspace{0.3cm}
\hspace{0cm}
   \includegraphics[width=6.5cm]{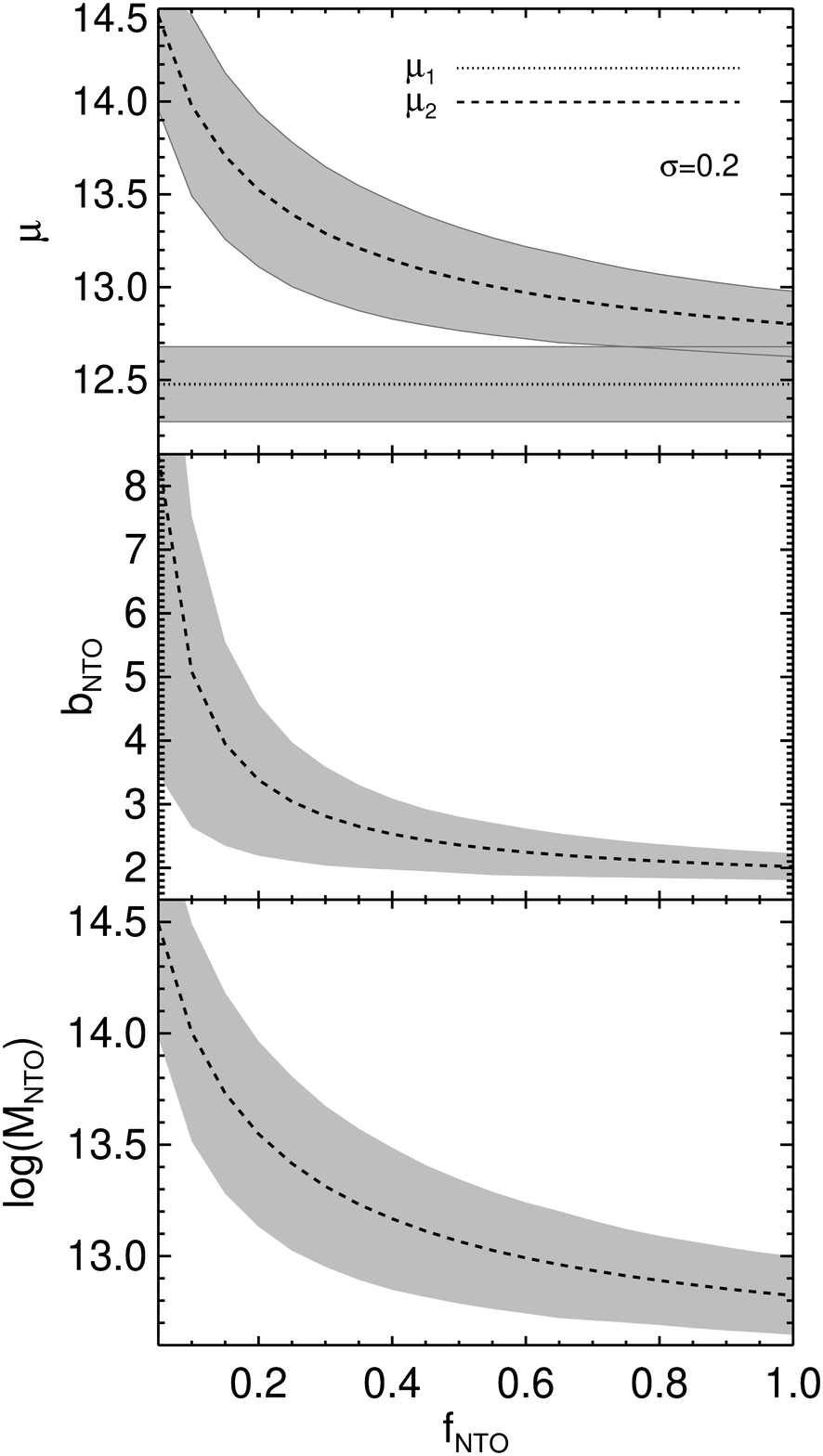}
    \vspace{0cm}
  \caption{The effect of the non-torus obscured fraction of the obscured subsample on the adopted $\mu_2$ and resulting non-torus obscured bias and halo mass.  Grey bands show Monte-Carlo errors as in Figure~\ref{fig:sig_trends}, and panels show the same properties.  At small $f_{\textrm{NTO}}$, the non-torus obscured objects must have very large halo masses and bias, in order to inflate the observed full obscured sample values.  As $f_{\textrm{NTO}}$ approaches unity, the values and errors of each parameter approach the observed values, since all objects are non-torus obscured.  The reality likely lies between these two scenarios (see section 5).  Note the somewhat sharp ``elbow'' in the $b_{\textrm{NTO}} - f_{\textrm{NTO}}$ relationship around $f_{\textrm{NTO}} \sim 0.2$, where small changes in the assumed $f_{\textrm{NTO}}$ can lead to large changes in the inferred $b_{\textrm{NTO}}$.\label{fig:fnto_trends}}
\vspace{0.2cm}
\end{figure}

Next we turn to the non-torus obscured sample behavior as a function of $f_{\textrm{NTO}}$, as shown in Figure~\ref{fig:fnto_trends}. The panels are the same as Figure~\ref{fig:sig_trends}, and the error ranges are estimated via the same resampling of the observed biases.  A value of $f_{\textrm{NTO}}=0$ implies that the entire obscured sample is torus obscured.  In our model, this value is only a limit and is not physically possible, as the assumption is that the torus obscured objects have the same bias as the unobscured objects, but the observed obscured bias is higher than that of the unobscured.  This is reflected in the rapidly increasing values of $\mu_2$, $b_{\textrm{NTO}}$, and $M_{\textrm{NTO}}$ with decreasing $f_{\textrm{NTO}}$.  At the opposite extreme, $f_{\textrm{NTO}}=1$, there is no dusty torus and all objects are non-torus obscured.  In this case, the non-torus obscured bias and mass is simply the observed bias and mass.  The reality is likely somewhere between these two extremes, as discussed in section 5.

We tabulate results at three distinct non-torus obscured fractions ($f_{\textrm{NTO}} = 0.25, 0.50, 0.75$) in Table~\ref{tbl:analytical}.  The top half shows results using the cosmology and matter power spectrum parameters consistent with \textsc{MultiDark}, and the bottom half shows the results using the parameters of D16.  Note that these are quite similar, but we include both for direct comparison of these analytical results with simulations in section 4 and with D16.

\begin{table*}
  \caption{Summary of analytical results.}
  \label{tbl:analytical}
  \begin{tabular}{lcccccccc}
  \hline
                                        &  &  $\mu_1$ ($M_{\odot}/h$)  & $\mu_2$ ($M_{\odot}/h$) & $b_{\textrm{NTO}}$ & $M_{\textrm{NTO}}$ ($M_{\odot}/h$)  &       C     &    $\theta_{\textrm{C,ST}}$ (deg) & $\theta_{\textrm{C,CT}}$ (deg) \\
  \hline
                                        &  &                            \multicolumn{7}{c}{\textsc{MultiDark} Parameters}                \\          
                                                                                      \cline{3-9}                \\       
$f_{\textrm{NTO}}=0.25$ &  &       12.48$\pm$0.20           &  13.39$\pm$0.39              &  3.04$\pm$0.93      &                 13.42$\pm$0.39                   &    0.33     &                       70.5                    &                  76.1                   \\
$f_{\textrm{NTO}}=0.50$ &  &       12.48$\pm$0.20           &  13.04$\pm$0.28              &  2.36$\pm$0.44      &                 13.07$\pm$0.28                   &    0.25     &                       75.5                    &                  79.7                    \\
$f_{\textrm{NTO}}=0.75$ &  &       12.48$\pm$0.20           &  12.89$\pm$0.21              &  2.13$\pm$0.29      &                 12.91$\pm$0.21                   &    0.14     &                       81.8                    &                  84.2                     \\
\\
                                      &  &                            \multicolumn{7}{c}{D16 Parameters}                \\          
                                                                                      \cline{3-9}                   \\   
$f_{\textrm{NTO}}=0.25$ &  &       12.60$\pm$0.19           &  13.46$\pm$0.37              &  3.08$\pm$0.93      &                 13.48$\pm$0.37                   &     0.33    &                       70.5                    &                  76.1                   \\
$f_{\textrm{NTO}}=0.50$ &  &       12.60$\pm$0.19           &  13.13$\pm$0.26              &  2.40$\pm$0.44      &                 13.15$\pm$0.26                   &     0.25    &                       75.5                    &                  79.7                    \\
$f_{\textrm{NTO}}=0.75$ &  &       12.60$\pm$0.19           &  12.98$\pm$0.20              &  2.17$\pm$0.29      &                 13.00$\pm$0.20                   &     0.14    &                       81.8                    &                  84.2                     \\
\hline
   \end{tabular}
   \\  
{
\raggedright    
 Summary of analytical results for three values of $f_{\textrm{NTO}}$.  The values of $\mu_1$ and $\mu_2$ are the means of the log-normal mass distributions of the unobscured/torus obscured and non-torus obscured samples, respectively, and $b_{\textrm{NTO}}$ and $M_{\textrm{NTO}}$ are the bias and ``typical'' halo mass of the non-torus obscured subsample. For our observed obscured fraction of 40\%, $C$ is the implied torus covering factor for each $f_{\textrm{NTO}}$ (Equation~\ref{eq:cf}, Figure~\ref{fig:cf}).  The angles $\theta_{\textrm{ST}}$ and $\theta_{\textrm{CT}}$ are the half-opening angles of the torus for a simple smooth torus and a clumpy torus, respectively, implied by these covering factors (see section 5.1).  The top half shows results using the same cosmology and power spectrum parameters as the \textsc{MultiDark} simulations, for direct comparison with our simulated results.  The bottom half shows results with the same cosmology and power spectrum parameters as D16, for a direct comparison with real-world measurements. \\
 }
\end{table*}

\section{SIMULATED QUASAR AUTOCORRELATIONS}
In the previous section we used purely analytical methods to determine the parameters of interest, namely $b_{\textrm{NTO}}$ and $M_{\textrm{NTO}}$.  Here, we turn to cosmological simulations in order to illustrate what an angular clustering measurement of these samples would look like.  This also provides an avenue to determine $b_{\textrm{NTO}}$ under different assumptions about the halo mass distribution that are less well-behaved and may not be analyzed analytically.

In addition, we used Monte-Carlo methods to estimate errors in the previous section, based on the errors of our observable parameters.  However, this does not necessarily track other, more subtle, potential sources of error.  For example, there are several parameters with potentially significant covariance, such as halo mass and clustering on different scales.  Working with simulated data provides a simple way to track such complicated issues, and also can help verify that the errors derived for our observational results are sensible.  

\subsection{Cosmological $N$-body simulations}
There are a number of cosmological simulations available to the community, which have steadily improved in spatial and mass resolution, as well as physical complexity --- for example, \textsc{Millennium} and \textsc{Millennium-II} \citep{2005Natur.435..629S, 2009MNRAS.398.1150B}, \textsc{Bolshoi} \citep{2011ApJ...740..102K}, \textsc{MultiDark} \citep{2012MNRAS.423.3018P}, and \textsc{Illustris} \citep{2014MNRAS.444.1518V, 2014MNRAS.445..175G, 2015MNRAS.452..575S}.  In this analysis we work with data from the \textsc{MultiDark} MDR1 simulation catalogue. We choose this simulation primarily because it has the largest volume, which allows us to easily build a simulated area on the same order as our observational region, and it has a cosmology similar to D16.  We verify however that applying our method below to a different simulation (\textsc{Millennium-II}) provides similar results, once differences in cosmology are accounted for.

The \textsc{MultiDark} MDR1 simulation\footnote{\url{https://www.cosmosim.org/cms/simulations/multidark-project/mdr1/}} has a box size of 1 Gpc/h with a mass resolution of $8.72 \times 10^9$ M$_{\odot}/h$.  The simulation begins at $z=65$, and the catalogue contains 85 snapshots at various redshifts from 65 to zero.  Our interest is at $z=1$, which corresponds to snapshot 52.  We use a simple SQL query of the MDR1 Bound Density Maximum (BDMV) table, which defines haloes based on their mass over density relative to the background,  to extract the positions of haloes in the mass range $11.6-14.5$ $\log($M$_{\odot}/h)$. This range is wide enough to encompass all of the masses needed when resampling our observed biases to estimate errors, using a log-normal distribution with $\sigma=0.2$:
\newline

\noindent \texttt{SELECT \\
snapnum, hostFlag, x, y, z, Mvir \\
FROM MDR1.BDMV WHERE snapnum = 52 AND \\
Mvir BETWEEN 3.98108e+11 AND 3.16228+14} \\  
\newline

\noindent This query selects 7 722 250 haloes that define the set from which we draw simulated samples.  

\subsection{Building simulated samples}
\begin{figure*}
\centering
\vspace{0.3cm}
\hspace{0cm}
   \includegraphics[width=17cm]{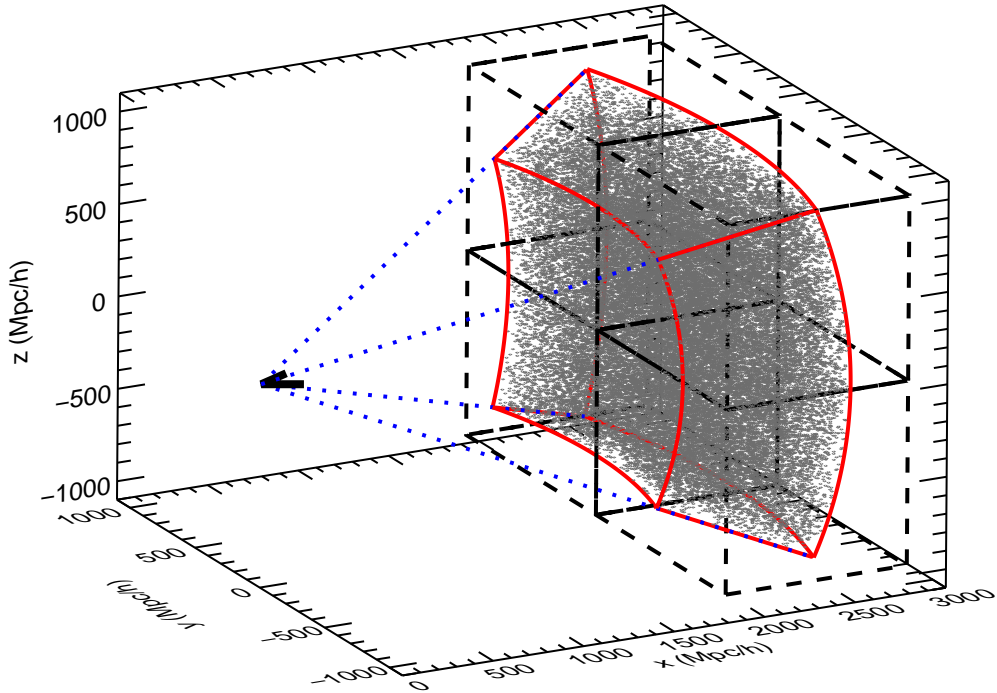}
    \vspace{0cm}
  \caption{The geometry used to convert the \textsc{MultiDark} simulation data into observations.  In order to increase the simulated area, we use four simulation boxes stacked in a 2-by-2 grid (black dashed boxes).  These are placed such that their centers are at the comoving distance corresponding to $z=1$ (in Mpc/$h$).  The observer is at the origin.  Using the full boxes would cause fluctuations in observed number densities as a function of position due to projection effects.  Therefore, the box is trimmed such that any line of sight sees through a constant thickness (red volume).  A random sample of haloes with the desired mass distribution is sampled (without replacement) from each box, such that the total number density matches the observed number density for the sample under consideration (gray points).  This is repeated 50 times (with replacement) to build 50 mock samples for each subsample and parameter set.  Finally, the $(x,y,z)$ positions are converted into spherical coordinates $(\chi,RA,Dec)$, where $\chi$ is the comoving distance.\label{fig:sim_boxes}}
\vspace{0.2cm}
\end{figure*}

We aim to build simulated samples that are similar to the real observational data of D16 in terms of area ($\sim$3000 deg$^2$) and number density ($\sim$30 and $\sim$20 deg$^{-2}$ for unobscured and obscured quasars, respectively).  After applying all of our cuts (described below), a single box would result in a simulated area of only $\sim$390 deg$^2$.  Given the periodic boundaries of the simulation boxes, we instead build a 2-by-2 grid of four boxes, each with their center placed at 2355 Mpc/h (the comoving distance $\chi$ to $z=1$) from the origin, as in Figure~\ref{fig:sim_boxes}.  This results in a final useable area of $1570$ deg$^2$, a large enough increase that concerns regarding edge effects and the finite area used for the clustering measurement (e.g. the integral constraint) are mitigated.  This grid of simulation boxes will introduce periodicities on the scale of a single box, roughly 20$^{\circ}$, which is far larger than the scales of interest for the autocorrelation measurements ($<1^{\circ}$, see below), and so does not introduce any bias.

Because of projection effects, we cannot use the full volume of each simulation box without having to account for density variations, as different lines of sight will traverse different path lengths through the box.  The simplest solution is to simply carve out a volume from the simulated set of boxes such that all path lengths are equal, as shown by the red region in Figure~\ref{fig:sim_boxes}.  The measurements that define this region are the maximum angular size of the back of the box (from the observers position at the origin), the distance to the front of the box along a line of sight to a back corner (where the blue dashed lines intersect the front of the box in Figure~\ref{fig:sim_boxes}), and the distance along the $x$-axis to the back of the box.

At the halo masses of interest ($\sim$10$^{13}$ M$_{\odot}/h$), the satellite occupation of quasars is negligible, meaning that the occupation function is dominated by quasars in central galaxies --- only at mass $\gtrsim14.5$ M$_{\odot}/h$ do haloes have on average one satellite quasar \citep[e.g.][]{2012ApJ...755...30R}.  This fact is also supported by a comparison of quasar clustering in perpendicular directions, which illustrates that quasars do not show large peculiar motions within haloes, and thus are dominated by central galaxies \citep[e.g.][]{2011ApJ...741...15S}.  This implies that we can simply convert the halo positions from the simulations to quasar positions, without populating the haloes based on kinematic properties, once we've selected the haloes that host quasars.  In order to select haloes from a given simulation box, we assign each a probability based on mass consistent with a log-normal distribution with $\sigma=0.2$ and mean $\mu$ as predicted by the analytical process in section 3.1.  We then sample $N$ haloes without replacement from a single box, and repeat this four times (once for each box), simply to reduce computational time by parallelizing the sample selection.  Histograms of the halo masses selected in this way are shown in Figure~\ref{fig:dndm}.  Note that while there will not be duplicate haloes in a single box, an individual halo can appear multiple times in a simulated observation by being selected in more than one box.  The value of $N$ is tuned so that the final sample has the appropriate number density when the used volume is projected onto the simulated sky.  We repeat this resampling 50 times, giving 50 independent samples of each of the three (unobscured, total obscured, non-torus obscured) mass distributions predicted for $f_{\textrm{NTO}}=0.25, 0.5$, and 0.75.  Note that the occupation fraction of halos (i.e.\ the fraction of selected halos at a given mass relative to the total number of available halos at that mass) is never higher than $\sim$15\%, meaning that there are enough halos available at each mass to select sufficiently independent samples despite the rapidly dropping halo mass function (see also the discussion on errors below). The Cartesian $(x,y,z)$ coordinates of each halo are then converted to spherical $(\chi,RA,Dec)$ values, where $\chi$ is the comoving distance.

\subsection{Measuring the bias of simulated samples}
To measure the bias of our simulated samples, we follow the same procedures as D16 (see their section 3 and the linked code libraries for full details).  We measure the angular autocorrelation $\omega(\theta)$ using the \citet{1993ApJ...412...64L} estimator, which compares data pair counts in annuli of increasing radii with those of a random distribution:
\begin{equation}
   \omega(\theta) = \frac{DD - 2DR + RR}{RR}.
   \label{eq:LS}
\end{equation}
The random sample is always at least 10 times larger than the simulated data set, and is generated using the \textsc{Mangle} utility \textsc{Ransack} \citep{2004MNRAS.349..115H, 2008MNRAS.387.1391S}.  We perform this calculation for each of the 50 random samplings individually, and adopt the mean at each $\theta$ as the final $\omega(\theta)$.

\begin{figure}
\centering
\vspace{0.3cm}
\hspace{0cm}
   \includegraphics[width=8.5cm]{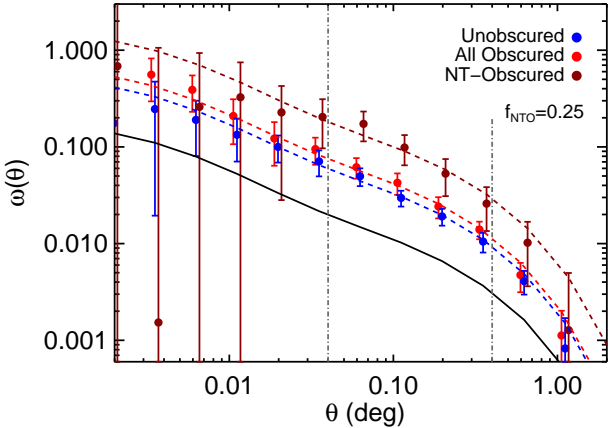}
    \vspace{0cm}
  \caption{The points show a simulated angular cross-correlation measurement, using haloes sampled from the \textsc{MultiDark} simulations, and the solid black line is the dark matter autocorrelation (see section 4.3 and Equation~\ref{eq:omega_mod}).  Haloes for each sample are selected to have (a linear combination of) log-normal mass distributions with standard deviation 0.2 and means predicted from the analytical analysis, for $f_{\textrm{NTO}} =0.25$ (see Table~\ref{tbl:analytical}). Halo positions are converted into $RA$ and $Dec$ as shown in Figure~\ref{fig:sim_boxes}, and the measurement is made as with the real data in D16, including the fitting range marked by the grey lines.  All simulated results are listed in Table~\ref{tbl:sims}.\label{fig:sim_clustering}}
\vspace{0.2cm}
\end{figure}

We utilize the 50 random samplings from the full set of simulated haloes in order to bootstrap the errors on $\omega(\theta)$ by generating the covariance matrix:
\begin{equation}
\textrm{\textsf{\textbf{C}}}_{ij} = \frac{1}{N-1} \sum_{L=1}^{N} [\omega_L(\theta_i) - \omega(\theta_i)] ~\times [\omega_L(\theta_j) - \omega(\theta_j)],
\label{eq:jack}
\end{equation}
where $\omega_L$ is the autocorrelation for a given sampling, and $i$ and $j$ are bins in angular scale.  The square-root of the diagonal elements of the covariance matrix are adopted as the 1-$\sigma$ errors.  An example of a simulated autocorrelation measurement with errors, for three samples (unobscured, complete obscured, and non-torus obscured) is shown in Figure~\ref{fig:sim_clustering}.

We use the standard Limber approximation (valid in the flat cosmology and small angular scales $\theta << 1$ radian probed here, and recast into distance rather than redshift space) to generate a model dark matter autocorrelation with bias unity:
\begin{equation}
 \omega_{dm} (\theta) = \pi \int_{\chi=0}^{\infty} \int_{k=0}^{\infty} \frac{\Delta^2 (k)}{k} J_0 [k \theta \chi] \left( \frac{dN}{d\chi} \right)^2 \frac{dk}{k} d\chi,
\label{eq:omega_mod}
\end{equation}
where $\chi$ is the comoving distance, $\Delta(k)^2$ is the dimensionless non-linear matter power spectrum at $z=1$ (generated with \textsc{CAMB}\footnote{\textit{Code for Anisotropies in the Microwave Background} (\url{http://lambda.gsfc.nasa.gov/toolbox/tb_camb_ov.cfm})}), and $J_0$ is the zeroth-order Bessel Function of the first kind.  The distance distribution of simulated haloes $dN/d\chi$ is generated from a spline fit to the radial distances $\chi$ to each halo.  This formalism properly handles the projection of the correlation function from three to two dimensions while assuming no evolution of the matter power spectrum across the box, which is made up of haloes all at $z=1$.  This model is shown as the black solid line in Figure~\ref{fig:sim_clustering}, and because of the geometry (Figure~\ref{fig:sim_boxes}) is the same for all samples.

The model $\omega_{\textrm{dm}}(\theta)$ is related to the measured $\omega(\theta)$ through the bias $b$, as $\omega(\theta) = b^2 \omega_{\textrm{dm}}(\theta)$.  We fit the model autocorrelation to the data using the covariance matrix and a chi-squared minimization:
\begin{equation}
\chi^2 = \sum_{i,j} [\omega(\theta_i) - \omega_{\textrm{dm}}(\theta_i)] \textsf{\textbf{C}}_{i,j}^{-1} [\omega(\theta_j) - \omega_{\textrm{dm}}(\theta_j)].
\end{equation}
We use the same fitting range as D16, $0.04^{\circ} < \theta < 0.4^{\circ}$ (though we note that widening this range to $0.01^{\circ} < \theta < 1.0^{\circ}$ has no significant impact on the results as most of the fitting power is in the smaller range), and errors on the bias are adopted where $\Delta \chi^2=1$.

\subsection{Results}
The simulated bias results are listed in Table~\ref{tbl:sims} for the same $f_{\textrm{NTO}}$ values as the analytical results in Table~\ref{tbl:analytical}.  The unobscured sample is not affected by changing $f_{\textrm{NTO}}$, and so these results are the same in each case, and are quite consistent with our observed and analytical predictions.  The full obscured sample bias is also consistent with our observed analytical predictions, always falling at $b_{\textrm{obsc}} \approx 2$.  

The simulated autocorrelations, once the mean over many random samplings of haloes is taken, reflect the shape predicted by the model very well (Figure~\ref{fig:sim_clustering}).  On small angular scales, the number counts of non-torus obscured sources is quite low and the errors become increasingly large, but on the scales of interest where we fit the bias the autocorrelation is well-behaved.  These results provide promise for more detailed modeling of these measurements in the future.

We find that in general, the errors on the bias in our simulated measurements are a factor of $\sim$2 smaller than the errors on the observational measurements from D16.  Given the fact that we have perfect control of our simulated samples, and they are not subject to observational effects, this is not too surprising.  The observed samples likely contain some amount of contamination from non-quasars \citep[on the order of a few to 10\%,][]{2012ApJ...753...30S}, as well as potentially highly correlated observational noise from instrumental effects or the sky (which will affect the ground-based optical data used for the unobscured/obscured split).  In any case, the simulated errors behave primarily like Poisson noise, as expected, in the sense that the relative errors scale roughly with the square root of the relative number densities.  This is also reflected in the increasing errors on the non-torus obscured bias and mass as $f_{\textrm{NTO}}$, and therefore the number of non-torus obscured sources, decreases.  The fact that the errors are Poisson in nature is further evidence that the random samplings are independent and not biased by a high occupation fractions at high mass. These simulated errors imply that 1) the errors derived for our real data are sensible and 2) if we were to, in the future, be able to make a clustering measurement of non-torus obscured quasars, the errors would limit the reliability of the measurements to scales $\gtrsim$0.02 degrees (for $f_{\textrm{NTO}}=0.25$).  This would make it difficult to perform a full HOD analysis including the ``one-halo'' term.

\begin{table*}
  \caption{Summary of simulated results.}
  \label{tbl:sims}
  \begin{tabular}{lccccccccc}
  \hline
                                 &  &  \multicolumn{2}{c}{$f_{\textrm{NTO}}=0.25$}    & & \multicolumn{2}{c}{$f_{\textrm{NTO}}=0.5$}      & & \multicolumn{2}{c}{$f_{\textrm{NTO}}=0.74$}                \\
                                                 \cline{3-4}                                                                       \cline{6-7}                                                      \cline{9-10}
			        & &               $b$            &     $M$ ($M_{\odot}/h$)    & &            $b$             &     $M$ ($M_{\odot}/h$)     & &           $b$               &       $M$ ($M_{\odot}/h$)           \\    
                                                                                      \cline{3-10}            \\                                                                                              
\vspace{0.1cm}
Unobscured              &  &   1.73$\pm$0.09    &   12.55$_{0.11}^{0.10}$    & &   1.73$\pm$0.09    &     12.55$_{0.11}^{0.10}$   & &   1.73$\pm$0.09     &     12.55$_{0.11}^{0.10}$             \\
\vspace{0.1cm}
Obscured                  &  &   1.98$\pm$0.12    &   12.79$_{0.10}^{0.09}$    & &   2.02$\pm$0.11    &    12.83$_{0.10}^{0.09}$    & &   2.04$\pm$0.13     &     12.84$_{0.11}^{0.10}$             \\
Non-torus obscured  &  &   2.97$\pm$0.29    &   13.39$_{0.14}^{0.12}$    & &   2.34$\pm$0.17    &    13.06$_{0.12}^{0.10}$    & &   2.12$\pm$0.13     &     12.90$_{0.10}^{0.09}$             \\
\hline
   \end{tabular}
   \\  
{
\raggedright    
The bias and inferred typical halo masses measured from the simulated samples, for three different values of the non-torus obscured fraction $f_{\textrm{NTO}}$.  The results from the first column ($f_{\textrm{NTO}}=0.25$) are shown in Figure~\ref{fig:sim_clustering}.   \\
 }
\end{table*}

\section{DISCUSSION}
\subsection{The value of $f_{\textrm{NTO}}$}
Under the assumption that torus obscured quasars are intrinsically the same as unobscured quasars, a core component of unification-by-orientation models, they will share the same bias and therefore host halo mass.  Therefore, any difference in these measured values implies that some fraction of the obscured sample must be obscured by some other material, whether it be nuclear dust in some other configuration due to evolution of the torus, dust in the narrow-line region, or some other, potentially galactic scale, obscurer.  It also implies that the measured overall obscured bias is in fact a lower limit on the bias of these sources.

To this point we have presented our results simply as a function of $f_{\textrm{NTO}}$, while making no assumptions about a reasonable expectation for its value.  The fact that the obscured sample has a higher bias than the unobscured sample implies that it must be greater than zero, as shown by our modeling where the bias approaches infinity as $f_{\textrm{NTO}}$ approaches zero.  Here we explore some potential ways to place reasonable limits on $f_{\textrm{NTO}}$, and therefore $b_{\textrm{NTO}}$.

\subsubsection{Torus Models}
There have been many attempts to model the dusty torus in quasars, from a simple smooth ``doughnut'' \citep[with properties that may depend on other quasar parameters like luminosity, e.g.][]{1988ApJ...329..702K, 1991MNRAS.252..586L}, to ``clumpy'' tori \citep[e.g.][]{2006A&A...452..459H, 2008ApJ...685..147N, 2008ApJ...685..160N, 2012MNRAS.420.2756S}, which may be the outer edge of an accretion disk wind \citep[e.g.][]{2005ApJ...631..689E, 2012ApJ...749...32K}.  Most current analyses of IR spectra and SEDs of quasars favor a clumpy, rather than smooth, torus \citep[e.g.][]{2011MNRAS.414.1082M, Comastri:2014tu, 2015arXiv151107096H, 2015arXiv151103503M, 2015MNRAS.451.2991G}.  

Torus models alone can place limits on the torus opening angle, which in turn provides the torus covering factor and $f_{\textrm{NTO}}$.  For example, recent numerical simulation models \citep[][]{Dorodnitsyn:2015wz} show that the column density of  obscuring material (in sources with a high Eddington fraction, as is likely the case in our sample) rises rapidly between 70-80$^{\circ}$ from the symmetry axis.  These models estimate that the Compton thick portion of the torus begins at an angle of $72-75^{\circ}$, independent of luminosity \citep{2012ApJ...761...70D}.  

Many works have used torus models to fit observed quasar spectra and SEDs.  For example, \citet{Deo:2011p1917} use the clumpy torus models of \citet{2008ApJ...685..160N} in order to model the SEDs of quasars at $z\sim2$ and find a typical value of the half width of the torus for their fits is $\sim$15$^{\circ}$, implying a $\theta_C$ of $\sim$75$^{\circ}$.   This is consistent with the numerical models above.

\begin{figure}
\centering
\vspace{0.3cm}
\hspace{0cm}
   \includegraphics[width=7.5cm]{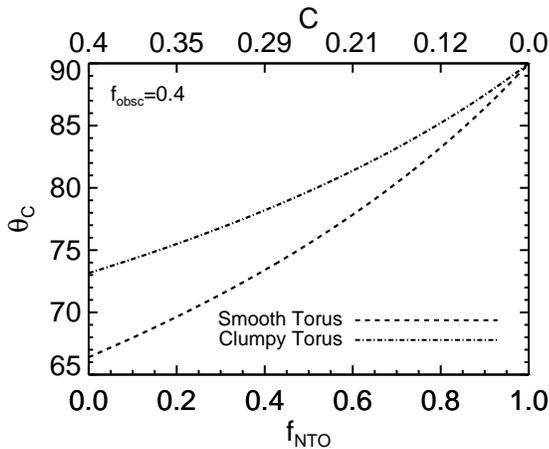}
    \vspace{0cm}
  \caption{For a given observed obscured fraction ($f_{\textrm{obsc}}=0.4$ for our sample), the covering factor of the torus (or simply nuclear obscuring dust) $C$ predicts a non-torus obscured fraction $f_{\textrm{NTO}}$.  For a given torus model, these can be related to the torus half opening angle $\theta_C$.  We show these relationships for a simple smooth torus with an abrupt edge (dashed line) and for a clumpy torus \citep[dot-dashed line, using the model of][with $m=2$ and $N_0=5$]{2008ApJ...685..160N}.  The observed $f_{\textrm{obsc}}$ places a lower limit on $\theta_C$ (upper limit on $C$), but if $\theta_c$ is larger than this (see section 5.1), then $f_{\textrm{NTO}}$ must increase to keep the observed obscured fraction constant.\label{fig:theta}}
\vspace{0.2cm}
\end{figure}

Given these potential constraints on the torus opening angle $\theta_C$, we can convert these to $f_{\textrm{NTO}}$ for our sample.  In Figure~\ref{fig:theta}, we show how, for our observed value of $f_{\textrm{obsc}} =0.4$, $f_{\textrm{NTO}}$ (and thus $C$) depends on $\theta_C$ for two different torus models.  For the simple, smooth torus case (dashed line), $\theta_C$ is the opening angle to the abrupt edge of the torus.  For the clumpy torus case (dot-dashed line), we adopt the ``soft edge'' torus model of \citet{2008ApJ...685..160N}, where the torus has a Gaussian distribution of clouds along the line of sight angle $\beta$: $N(\beta)=N_0 e^{-|\beta/(1-\theta_C)|^m}$, where $\theta_C$ represents the angle to the half-width of the Gaussian cloud distribution.  Here we use moderate values in their parameter space of $N_0=5$ and $m=2$.  The torus covering factor $C$ in this model is:
\begin{equation}
C = 1 - \int_0^{\pi/2} e^{-N(\beta)} \cos \beta d\beta.
\end{equation}
We include values of $\theta_C$ for the two models at various $f_{\textrm{NTO}}$ in Table~\ref{tbl:analytical}.

For the values discussed above, where $\theta_C$ is roughly 75$^{\circ}$, this implies $f_{\textrm{NTO}} \sim 0.25$ and $f_{\textrm{NTO}} \sim 0.45$ in the clumpy and smooth torus models, respectively.  Since clumpy models are generally better at reproducing the observed properties of quasars, it seems that our results for $f_{\textrm{NTO}} = 0.25$ represent the most realistic values, given current knowledge.  Interestingly, this is right near the ``elbow'' in the $b_{\textrm{NTO}} - f_{\textrm{NTO}}$ plane (see Figure~\ref{fig:fnto_trends}), where small changes can strongly impact the inferred non-torus obscured bias.

\subsubsection{Observational constraints}
The observed fraction of obscured sources, $f_{\textrm{obsc}}$, is directly related to the \emph{total} dust covering factor and the total obscured fraction.  The obscured fraction has been studied by many groups, often via X-rays, with general agreement that $f_{\textrm{obsc}}$ depends on $L_{\textrm{bol}}$ (and may have a dependence on redshift), and general consistency with our $f_{\textrm{obsc}}=0.4$ at $L_{\textrm{bol}} \sim 10^{46}$ ergs/s \citep[e.g.][]{2003ApJ...598..886U, 2005ApJ...635..864L, 2008ApJ...679..140T, 2014MNRAS.437.3550M, 2015ARA&A..53..365N}.  Note however, that there are often many biases and caveats to these population studies, due to e.g.\ the uncertain fraction of Compton-thick AGN that may be missed completely.  However, the difficulty is that in order to determine $f_{\textrm{NTO}}$, we need to know the \emph{torus} covering factor as well, so that we can separate the two components.  While torus models can provide this as discussed above, some work has been done to constrain this observationally.

\citet{2008ApJ...679..140T} approached determining the torus covering factor in luminous unobscured quasars by examining the ratio of $L_{IR}$, assumed to be quasar light reprocessed by the torus, to $L_{\textrm{bol}}$, but still relied somewhat on torus models to convert these into observed obscured fractions.  At the luminosity of our sample, their prediction is that the torus obscured fraction should be $\sim$0.5.  This is larger than their overall observed obscured fraction (and ours as well), which may be (at least partially) due to missing a large population of Compton thick sources.  It is difficult to say how these missing sources will affect $f_{\textrm{NTO}}$, as Compton thick obscuration can happen in the torus as well as in the host galaxy.  In any case, it would seem that these observations imply a small $f_{\textrm{NTO}}$, since they predict a fairly large torus covering factor.

Though optical spectroscopic follow-up of statistically complete mid-IR selected samples is not available, current data can provide some qualitative insight.  For example, \citet{2013ApJS..208...24L} find that 22\% of mid-IR selected quasars do not have clear AGN signatures in their emission lines, though many still have radio and/or X-ray properties consistent with AGN activity. \citet{2014ApJ...795..124H} find that $\sim$12\% of \wise\ selected obscured quasars do not have strong emission features, and potentially an additional $\sim$10\% that have non-AGN dominated emission lines.  These studies suggest that NTO quasars make up on the order of $\sim$20\% of the obscured population.

Another way to explore the nature of the obscuring material is to analyze IR spectra, which can shed light on the dust properties and location via e.g.\ silicate absorption features.  This was exploited for example by \citet{2012ApJ...755....5G}, who used the \textsc{Si} $\lambda \sim 9.7$ $\mu$m feature along with galaxy inclinations to show that a significant population of obscured sources is affected by large-scale dust.  \citet{2008ApJ...675..960P} studied the obscuration in luminous quasars at $z>1$, and find that a torus plus cold absorber is needed to explain their full range of IR spectra.  Their observed $f_{\textrm{obsc}}$ is much higher than ours, at $\sim$0.6, and their value of $f_{\textrm{NTO}}$ would be $\sim$0.5.  This discrepancy may be due to the possible redshift dependence of the obscured fraction, which increases at higher $z$ for high luminosity quasars \citep[e.g.][]{2014MNRAS.437.3550M}, and the fact that the \citet{2008ApJ...675..960P} mean redshift is higher than ours.  It could also be evidence that $f_{\textrm{NTO}}$ is higher than the 0.25 suggested above.  While current samples of objects with high-quality IR spectra are small, a future possibility would be to explore the clustering and bias of sources with different types of obscuration based on their IR spectral features.  This could shed additional light on the haloes of the non-torus obscured population.

\subsection{The hosts of non-torus obscured quasars}
\subsubsection{Relative space densities \& lifetimes}
Assuming similar bolometric luminosity and redshift distributions for the torus-obscured and non-torus obscured samples, which we argued in section 2 is reasonable, we can predict the non-torus obscured quasar space density as a function of $f_{\textrm{NTO}}$.  Using the bolometric luminosity function of \citet{2007ApJ...654..731H} at $z=1$, we find for $L_{\textrm{bol}} \sim 10^{46}$ \citep[characteristic of mid-IR selected quasars,][]{2011ApJ...731..117H, 2014ApJ...795..124H} an overall space density of quasars of $\sim$2$\times 10^{-5}$ Mpc$^{-3}$.  The space density of non-torus obscured quasars, then, is this overall density times the non-torus obscured fraction of \textit{all} quasars (not just the non-torus obscured fraction of the obscured quasars, which we have been labeling as $f_{\textrm{NTO}}$).  This fraction is given for our sample by $f_{\textrm{NTO,tot}} = f_{\textrm{NTO}} \times f_{\textrm{obsc}}=f_{\textrm{NTO}} \times 0.4$.

\begin{figure}
\centering
\vspace{0.3cm}
\hspace{0cm}
   \includegraphics[width=7.5cm]{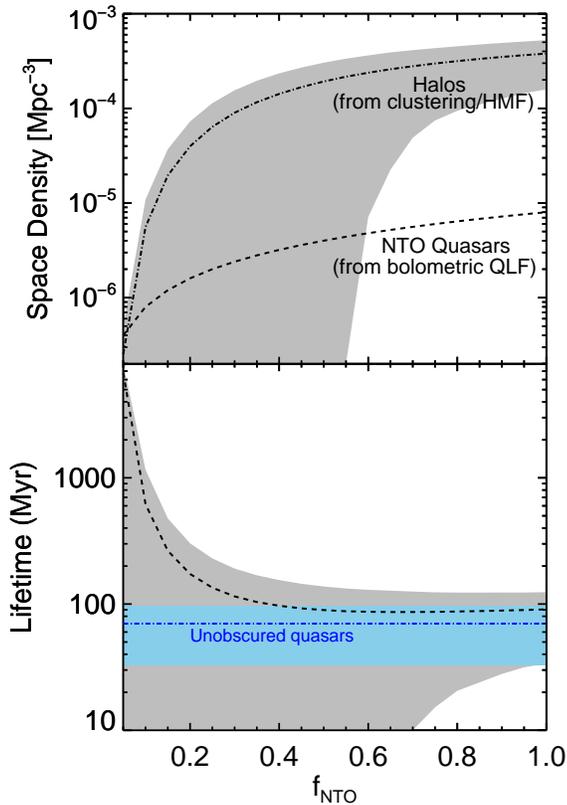}
    \vspace{0cm}
  \caption{\emph{Top:} The estimated space densities of the non-torus obscured quasars and potential host haloes with $M_{\textrm{NTO}}$ (from clustering and the halo mass function of \citealt{2010ApJ...724..878T}), as a function of $f_{\textrm{NTO}}$.  The non-torus obscured space density is based on the total non-torus obscured fraction ($f_{\textrm{NTO}} \times f_{\textrm{obsc}}$, where $f_{\textrm{obsc}}=0.4$) and the bolometric quasar luminosity function of \citet{2007ApJ...654..731H} at $z=1$.  \emph{Bottom:} The inferred non-torus obscured quasar lifetime based on abundance matching and assuming the bulk of activity occurs over $0.5 < z < 1.5$. The range of lifetimes for \wise\ unobscured quasars from D16 is also shown in blue for comparison.\label{fig:dense}}
\vspace{0.2cm}
\end{figure}

Similarly, we can find the predicted space density of haloes with mass $M_{\textrm{NTO}}$ for each $f_{\textrm{NTO}}$ using the halo mass function of \citet{2010ApJ...724..878T}.  Note that if the predicted halo space density is lower than that predicted for the non-torus obscured sources, this is another potential constraint on $f_{\textrm{NTO}}$, because each non-torus obscured quasar must have a halo to reside in. The densities are shown in the top portion of Figure~\ref{fig:dense}, along with error ranges based on the errors of $M_{\textrm{NTO}}$.  Though the errors are large, only extremely small values ($<$0.02) of $f_{\textrm{NTO}}$ are ruled out by these estimates.  However, up to $f_{\textrm{NTO}} \sim 0.55$, the halo densities are consistent with the non-torus obscured density, within the errors, suggesting that $f_{\textrm{NTO}}$ may be larger than the estimates from torus models.  Improved measurements of the obscured quasar bias, with larger samples, would improve these constraints.

We use the ratio of the non-torus obscured quasar density and halo density \citep[abundance matching;][]{1999ApJ...523...32C, 1999ApJ...520..437K, 2004MNRAS.353..189V, 2006ApJ...643...14S, 2010MNRAS.404.1111G}, along with the cosmic time from $0.5 < z < 1.5$, where the bulk of these sources lie, to estimate the average lifetime of the non-torus obscured quasars, as shown in the bottom half of Figure~\ref{fig:dense}.  For most values of $f_{\textrm{NTO}}$, the duty cycle of the non-torus obscured quasars is about the same as that of unobscured quasars ($\sim$100 Myr), and on the order of $\sim$1\% of the Hubble time. Only in the case where these sources are exceedingly rare ($f_{\textrm{NTO}} < 0.1$) is their lifetime estimated to be significantly larger.

\subsubsection{Cosmic evolution of non-torus obscured quasars}
If we adopt $f_{\textrm{NTO}}=0.25$, our best estimate based on torus models, this implies $b_{\textrm{NTO}}=3.08$ and $\log M_{\textrm{NTO}}=13.48$ M$_{\odot}/h$ at $z=1$.  Using the merger rate and mass assembly analysis of \citet{2010MNRAS.406.2267F}, we can extrapolate the halo masses of non-torus obscured quasars to earlier and later times in order to speculate on their progenitors and descendants \citep[e.g][]{2012MNRAS.421..284H}.

\begin{figure}
\centering
\vspace{0.3cm}
\hspace{0cm}
   \includegraphics[width=8cm]{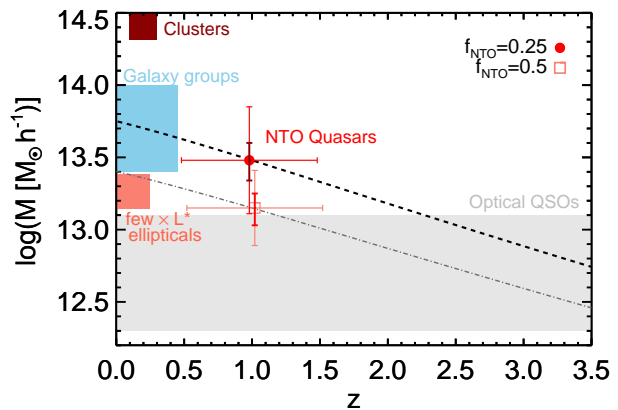}
    \vspace{0cm}
  \caption{Extrapolating the halo mass of non-torus obscured obscured quasars (red point) over cosmic time (dashed line), using the formalism of \citet{2010MNRAS.406.2267F} and assuming $f_{\textrm{NTO}}=0.25$.  The $x$-axis error bars indicate the redshift range of most of the total obscured sample, assumed to be the same for the non-torus obscured sample.  The two $y$-axis error bars show the errors from our analytical predictions (red), which are based on the observed bias errors, and the smaller errors from the covariance matrix of the simulations (dark red).  We also show the results for $f_{\textrm{NTO}} = 0.5$ for completeness, though we argued that this value was less likely based on torus models and observations in section 5.1.  Some representative regions for other classes of objects are shown for comparison (see section 5.2.2).  Progenitors of NTO quasars at $z=1$ could be quasars in the most massive haloes, and they end up as members of galaxy groups at $z=0$.\label{fig:growth}}
\vspace{0.2cm}
\end{figure}

Adopting the median growth rate of haloes as a function of mass and redshift (note that this is lower than the mean growth rate), we project the derived halo mass over $0< z<3.5$, shown in Figure~\ref{fig:growth} (we also show a similar analysis for $f_{\textrm{NTO}}=0.5$ and $M_{\textrm{NTO}}=13.15$ M$_{\odot}/h$ for comparison). 
Taking the bias measurements from an assortment of other works and converting these to halo masses in our cosmology and using our matter power spectrum parameters, we highlight general mass ranges at different stages of cosmic time for other object classes: galaxy clusters \citep{2009ApJ...692..265E}, galaxy pairs/groups \citep{2014MNRAS.444.2854W, 2015MNRAS.446.1356H}, massive ellipticals \citep{2011ApJ...736...59Z}, and unobscured quasars \citep[e.g.][D16]{2004MNRAS.355.1010P, 2005MNRAS.356..415C, 2007ApJ...654..115C, 2007ApJ...658...85M, 2008MNRAS.383..565D, 2009MNRAS.397.1862P, 2009ApJ...697.1634R, 2010ApJ...713..558K, 2012MNRAS.424..933W, 2013ApJ...778...98S, 2014MNRAS.442.3443D, 2015MNRAS.446.3492D, 2015MNRAS.453.2779E}. 

Based on these projections, it is possible that the non-torus obscured quasars often also went through an unobscured quasar phase at high redshift.  This scenario doesn't contradict, as it may initially seem, evolutionary scenarios that suggest that the obscured phase is an early part of the quasar duty cycle followed by a blowout and unobscured phase \citep[e.g.][]{2008ApJS..175..356H}.  \citet{2011ApJ...731..117H}, and later \citet{2014MNRAS.442.3443D} and D16, suggested that the higher halo masses of obscured quasars fit into such a model, with the early obscured phase representing a period in which the black hole is ``catching up'' to its final mass relative to its dark matter halo \citep{2010MNRAS.408L..95K}. The possibility that the progenitors of the non-torus obscured population at $z=1$ are unobscured quasars at $z\sim3$ suggests a cyclical or stochastic nature in which multiple halo mergers can ignite quasar activity in distinct bursts, each with an obscured and unobscured phase.  This suggests that the non-torus obscured quasars at $z=1$ will go through an unobscured phase at slightly lower redshift, but (given the number density of the torus-obscured population) these will be rare objects and make up the high halo mass tail of the unobscured population.  These objects will then go on to form (quiescent) galaxy groups in the local Universe, which tend to reside in large haloes of mass $\sim$10$^{13.7}$ M$_{\odot}/h$.

We note finally that, if the adopted $f_{\textrm{NTO}}=0.25$ is correct, then $M_{\textrm{NTO}}$, even with its large associated error, still falls outside of the typical mass range of optically selected, unobscured quasars.  While this was true of the full obscured population in early measurements \citep{2014MNRAS.442.3443D, 2015MNRAS.446.3492D}, the most recent values of D16 placed the obscured quasar bias and thus halo mass within the unobscured range.  Our analysis in this current paper suggests, instead, that the obscured population that is truly distinct from the unobscured population falls outside of the range of halo masses measured for unobscured quasars. This is because some fraction of the obscured population, in being torus-obscured, must share the range of host halo masses of unobscured quasars. This necessarily inflates the halo masses for the remaining (now-smaller-fraction-of) non-torus obscured quasars. Note that this interpretation depends on the exact value of $f_{\textrm{NTO}}$; higher values of $f_{\textrm{NTO}}$ will reduce the inferred mass differences between non-torus obscured quasars and unobscured quasars (Figure~\ref{fig:growth}).

\section{SUMMARY}
Samples of obscured quasars are likely a mix of objects intrinsically the same as unobscured quasars but seen through a dusty torus, and a distinct population of objects obscured by dust in a different distribution, possibly on large-scales and stirred up by galaxy mergers or interactions.  Therefore, the recently measured higher host dark matter halo masses of obscured quasars is likely diluted by torus-obscured objects, and only provides a lower limit on the typical halo masses of the non-torus obscured population. 

Making simple assumptions about the halo mass distributions of these populations, informed by recent work on the halo occupation distribution of unobscured quasars, we provide estimates of the bias and typical host halo masses of the non-torus obscured population.  We provide these first as a function of the non-torus obscured fraction, using both analytical methods and mock angular autocorrelation measurements of haloes drawn from cosmological $N$-body simulations.  Current torus models and observations indicate that a reasonable value for the non-torus obscured fraction is $\sim$25\% of the full obscured population, which implies a bias of $b_{\textrm{NTO}} \approx 3$ and $\log (M_{\textrm{NTO}}) \approx 3 \times 10^{13}$ M$_{\odot}/h$.  For comparison, recent measurements have found $b_{\textrm{unobscured}} = 1.72$ and $\log (M_{\textrm{unobscured}}) = 12.56$ M$_{\odot}/h$. 

Our simulated measurements and errors suggest that at these halo masses and number densities, an angular autocorrelation measurement of non-torus obscured quasars should be distinguishable from that of unobscured or torus-obscured quasars, at least on large scales ($>0.02^{\circ}$).  However, the implied low number density may make it difficult to accurately measure small-scale clustering and perform a full halo occupation distribution analysis, unless a cross-correlation with a larger tracer sample is utilized.  Regardless, isolating a non-torus obscured sample in large enough numbers to perform these measurements will be difficult.

The results of this analysis fit nicely within a merger-based scenario for luminous obscured quasars, while still retaining a population of sources that fit within a unification by orientation model.  Based on the projection of the estimated non-torus obscured halo masses to higher and lower redshifts, we suggest that at earlier times ($z \approx 3$) these objects go through a separate, unobscured quasar phase (possibly the second half of an earlier obscured phase).  At later times, they may go through another unobscured phase post-blowout, and make up a rare population of unobscured quasars in a high-halo mass tail of the full distribution.  Their masses at $z=1$ suggest that they may be the progenitors of quiescent galaxy group members in the local Universe.

The higher mass haloes of obscured quasars, along with direct evidence of obscuration on large scales separate from the dusty torus, imply the existence of a non-torus obscured population.  Analytical predictions such as those provided here are useful to further explore the nature of these sources, and continue to refine additional factors of the unification by orientation model that is so successful in explaining many, but not all, quasar observations.

\section*{Acknowledgements}
MAD, RCH, and ADM were partially supported by NASA through ADAP award NNX12AE38G.  MAD and RCH were also partially supported by the National Science Foundation through grant numbers 1211096 and 1515364.  MAD and ADM were also supported by NSF grant numbers 1211112 and 1515404.  RCH also acknowledges support from an Alfred P. Sloan Research Fellowship, and a Dartmouth Class of 1962 Faculty Fellowship.

\bibliography{/Users/Mike/Dropbox/full_library}

\begin{thebibliography}{106}
\expandafter\ifx\csname natexlab\endcsname\relax\def\natexlab#1{#1}\fi

\bibitem[{Antonucci(1993)}]{1993ARA&A..31..473A}
Antonucci R., 1993, ARA{\&}A, 31, 473

\bibitem[{Assef {et~al}\mbox{.}(2013)Assef, Stern, Kochanek, Blain, Brodwin,
  Brown, Donoso, Eisenhardt, Jannuzi, Jarrett, Stanford, Tsai, Wu, \&
  Yan}]{2013ApJ...772...26A}
Assef R.~J. {et~al.}, 2013, ApJ, 772, 26

\bibitem[{Berlind \& Weinberg(2002)}]{2002ApJ...575..587B}
Berlind A.~A., Weinberg D.~H., 2002, ApJ, 575, 587

\bibitem[{Berlind {et~al}\mbox{.}(2003)Berlind, Weinberg, Benson, Baugh, Cole,
  Dav{\'e}, Frenk, Jenkins, Katz, \& Lacey}]{2003ApJ...593....1B}
Berlind A.~A. {et~al.}, 2003, ApJ, 593, 1

\bibitem[{Bian \& Gu(2007)}]{2007ApJ...657..159B}
Bian W., Gu Q., 2007, ApJ, 657, 159

\bibitem[{Booth \& Schaye(2010)}]{2010MNRAS.405L...1B}
Booth C.~M., Schaye J., 2010, MNRAS, 405, L1

\bibitem[{Boylan-Kolchin {et~al}\mbox{.}(2009)Boylan-Kolchin, Springel, White,
  Jenkins, \& Lemson}]{2009MNRAS.398.1150B}
Boylan-Kolchin M., Springel V., White S. D.~M., Jenkins A., Lemson G., 2009,
  MNRAS, 398, 1150

\bibitem[{Brown {et~al}\mbox{.}(2008)Brown, Zheng, White, Dey, Jannuzi, Benson,
  Brand, Brodwin, \& Croton}]{2008ApJ...682..937B}
Brown M. J.~I. {et~al.}, 2008, ApJ, 682, 937

\bibitem[{Chatterjee {et~al}\mbox{.}(2012)Chatterjee, DeGraf, Richardson,
  Zheng, Nagai, \& Di~Matteo}]{2012MNRAS.419.2657C}
Chatterjee S., DeGraf C., Richardson J., Zheng Z., Nagai D., Di~Matteo T.,
  2012, MNRAS, 419, 2657

\bibitem[{Chatterjee {et~al}\mbox{.}(2013)Chatterjee, Nguyen, Myers, \&
  Zheng}]{2013ApJ...779..147C}
Chatterjee S., Nguyen M.~L., Myers A.~D., Zheng Z., 2013, ApJ, 779, 147

\bibitem[{Chen {et~al}\mbox{.}(2015)Chen, Hickox, Alberts, Harrison, Alexander,
  Assef, Brodwin, Brown, Del~Moro, Forman, Gorjian, Goulding, Hainline, Jones,
  Kochanek, Murray, Pope, Rovilos, \& Stern}]{2015ApJ...802...50C}
Chen C.-T.~J. {et~al.}, 2015, arXiv, 802, 50

\bibitem[{Coil {et~al}\mbox{.}(2007)Coil, Hennawi, Newman, Cooper, \&
  Davis}]{2007ApJ...654..115C}
Coil A.~L., Hennawi J.~F., Newman J.~A., Cooper M.~C., Davis M., 2007, ApJ,
  654, 115

\bibitem[{Col{\'\i}n {et~al}\mbox{.}(1999)Col{\'\i}n, Klypin, Kravtsov, \&
  Khokhlov}]{1999ApJ...523...32C}
Col{\'\i}n P., Klypin A.~A., Kravtsov A.~V., Khokhlov A.~M., 1999, ApJ, 523, 32

\bibitem[{Comastri {et~al}\mbox{.}(2014)Comastri, Harrison, Alexander,
  Ballantyne, Bauer, Boggs, Brandt, Brightman, Christensen, Craig, Del~Moro,
  Gandhi, Hailey, Koss, Lansbury, Luo, Madejski, Marinucci, Matt, Markwardt,
  Puccetti, Reynolds, Risaliti, Rivers, Stern, Walton, \&
  Zhang}]{Comastri:2014tu}
Comastri A. {et~al.}, 2014, arXiv

\bibitem[{Croom {et~al}\mbox{.}(2005)Croom, Boyle, Shanks, Smith, Miller,
  Outram, Loaring, Hoyle, \& da~{\^A}ngela}]{2005MNRAS.356..415C}
Croom S.~M. {et~al.}, 2005, MNRAS, 356, 415

\bibitem[{Croton(2009)}]{2009MNRAS.394.1109C}
Croton D.~J., 2009, MNRAS, 394, 1109

\bibitem[{da~{\^A}ngela {et~al}\mbox{.}(2008)da~{\^A}ngela, Shanks, Croom,
  Weilbacher, Brunner, Couch, Miller, Myers, Nichol, Pimbblet, de~Propris,
  Richards, Ross, Schneider, \& Wake}]{2008MNRAS.383..565D}
da~{\^A}ngela J. {et~al.}, 2008, MNRAS, 383, 565

\bibitem[{Deo {et~al}\mbox{.}(2011)Deo, Richards, Nikutta, Elitzur, Gallagher,
  Ivezi{\'c}, \& Hines}]{Deo:2011p1917}
Deo R.~P., Richards G.~T., Nikutta R., Elitzur M., Gallagher S.~C., Ivezi{\'c}
  {\v Z}., Hines D., 2011, ApJ, 729, 108

\bibitem[{DiPompeo, Hickox \& Myers(2016)DiPompeo, Hickox, \&
  Myers}]{2016MNRAS.456..924D}
DiPompeo M.~A., Hickox R.~C., Myers A.~D., 2016, MNRAS, 456, 924

\bibitem[{DiPompeo {et~al}\mbox{.}(2014)DiPompeo, Myers, Hickox, Geach, \&
  Hainline}]{2014MNRAS.442.3443D}
DiPompeo M.~A., Myers A.~D., Hickox R.~C., Geach J.~E., Hainline K.~N., 2014,
  MNRAS, 442, 3443

\bibitem[{DiPompeo {et~al}\mbox{.}(2015)DiPompeo, Myers, Hickox, Geach, Holder,
  Hainline, \& Hall}]{2015MNRAS.446.3492D}
DiPompeo M.~A., Myers A.~D., Hickox R.~C., Geach J.~E., Holder G., Hainline
  K.~N., Hall S.~W., 2015, MNRAS, 446, 3492

\bibitem[{Dorodnitsyn \& Kallman(2012)}]{2012ApJ...761...70D}
Dorodnitsyn A., Kallman T., 2012, ApJ, 761, 70

\bibitem[{Dorodnitsyn, Kallman \& Proga(2015)Dorodnitsyn, Kallman, \&
  Proga}]{Dorodnitsyn:2015wz}
Dorodnitsyn A., Kallman T., Proga D., 2015, arXiv

\bibitem[{Eftekharzadeh {et~al}\mbox{.}(2015)Eftekharzadeh, Myers, White,
  Weinberg, Schneider, Shen, Font-Ribera, Ross, P{\^a}ris, \&
  Streblyanska}]{2015MNRAS.453.2779E}
Eftekharzadeh S. {et~al.}, 2015, MNRAS, 453, 2779

\bibitem[{Estrada, Sefusatti \& Frieman(2009)Estrada, Sefusatti, \&
  Frieman}]{2009ApJ...692..265E}
Estrada J., Sefusatti E., Frieman J.~A., 2009, ApJ, 692, 265

\bibitem[{Everett(2005)}]{2005ApJ...631..689E}
Everett J.~E., 2005, ApJ, 631, 689

\bibitem[{Fakhouri, Ma \& Boylan-Kolchin(2010)Fakhouri, Ma, \&
  Boylan-Kolchin}]{2010MNRAS.406.2267F}
Fakhouri O., Ma C.-P., Boylan-Kolchin M., 2010, MNRAS, 406, 2267

\bibitem[{Gallagher {et~al}\mbox{.}(2015)Gallagher, Everett, Abado, \&
  Keating}]{2015MNRAS.451.2991G}
Gallagher S.~C., Everett J.~E., Abado M.~M., Keating S.~K., 2015, MNRAS, 451,
  2991

\bibitem[{Geach {et~al}\mbox{.}(2013)Geach, Hickox, Bleem, Brodwin, Holder,
  Aird, Benson, Bhattacharya, Carlstrom, Chang, Cho, Crawford, Crites, de~Haan,
  Dobbs, Dudley, George, Hainline, Halverson, Holzapfel, Hoover, Hou, Hrubes,
  Keisler, Knox, Lee, Leitch, Lueker, Luong-Van, Marrone, McMahon, Mehl, Meyer,
  Millea, Mohr, Montroy, Myers, Padin, Plagge, Pryke, Reichardt, Ruhl, Sayre,
  Schaffer, Shaw, Shirokoff, Spieler, Staniszewski, Stark, Story, van Engelen,
  Vanderlinde, Vieira, Williamson, \& Zahn}]{2013ApJ...776L..41G}
Geach J.~E. {et~al.}, 2013, ApJL, 776, L41

\bibitem[{Genel {et~al}\mbox{.}(2014)Genel, Vogelsberger, Springel, Sijacki,
  Nelson, Snyder, Rodriguez-Gomez, Torrey, \& Hernquist}]{2014MNRAS.445..175G}
Genel S. {et~al.}, 2014, MNRAS, 445, 175

\bibitem[{Goulding {et~al}\mbox{.}(2012)Goulding, Alexander, Bauer, Forman,
  Hickox, Jones, Mullaney, \& Trichas}]{2012ApJ...755....5G}
Goulding A.~D., Alexander D.~M., Bauer F.~E., Forman W.~R., Hickox R.~C., Jones
  C., Mullaney J.~R., Trichas M., 2012, ApJ, 755, 5

\bibitem[{Guo {et~al}\mbox{.}(2010)Guo, White, Li, \&
  Boylan-Kolchin}]{2010MNRAS.404.1111G}
Guo Q., White S., Li C., Boylan-Kolchin M., 2010, MNRAS, 404, 1111

\bibitem[{Hainline {et~al}\mbox{.}(2014)Hainline, Hickox, Carroll, Myers,
  DiPompeo, \& Trouille}]{2014ApJ...795..124H}
Hainline K.~N., Hickox R.~C., Carroll C.~M., Myers A.~D., DiPompeo M.~A.,
  Trouille L., 2014, ApJ, 795, 124

\bibitem[{Hamilton \& Tegmark(2004)}]{2004MNRAS.349..115H}
Hamilton A. J.~S., Tegmark M., 2004, MNRAS, 349, 115

\bibitem[{Han {et~al}\mbox{.}(2015)Han, Eke, Frenk, Mandelbaum, Norberg,
  Schneider, Peacock, Jing, Baldry, Bland-Hawthorn, Brough, Brown, Liske,
  Loveday, \& Robotham}]{2015MNRAS.446.1356H}
Han J. {et~al.}, 2015, MNRAS, 446, 1356

\bibitem[{Hao {et~al}\mbox{.}(2011)Hao, Elvis, Civano, \&
  Lawrence}]{2011ApJ...733..108H}
Hao H., Elvis M., Civano F., Lawrence A., 2011, ApJ, 733, 108

\bibitem[{He, Liu \& Zhang(2015)He, Liu, \& Zhang}]{2015arXiv151107096H}
He J.-J., Liu Y., Zhang S.-N., 2015, arXiv, 7096

\bibitem[{Hickox {et~al}\mbox{.}(2007)Hickox, Jones, Forman, Murray, Brodwin,
  Brown, Eisenhardt, Stern, Kochanek, Eisenstein, Cool, Jannuzi, Dey, Brand,
  Gorjian, \& Caldwell}]{2007ApJ...671.1365H}
Hickox R.~C. {et~al.}, 2007, ApJ, 671, 1365

\bibitem[{Hickox {et~al}\mbox{.}(2011)Hickox, Myers, Brodwin, Alexander,
  Forman, Jones, Murray, Brown, Cool, Kochanek, Dey, Jannuzi, Eisenstein,
  Assef, Eisenhardt, Gorjian, Stern, Le~Floc'h, Caldwell, Goulding, \&
  Mullaney}]{2011ApJ...731..117H}
Hickox R.~C. {et~al.}, 2011, ApJ, 731, 117

\bibitem[{Hickox {et~al}\mbox{.}(2012)Hickox, Wardlow, Smail, Myers, Alexander,
  Swinbank, Danielson, Stott, Chapman, Coppin, Dunlop, Gawiser, Lutz, van~der
  Werf, \& Wei{\ss}}]{2012MNRAS.421..284H}
Hickox R.~C. {et~al.}, 2012, MNRAS, 421, 284

\bibitem[{H{\"o}nig {et~al}\mbox{.}(2006)H{\"o}nig, Beckert, Ohnaka, \&
  Weigelt}]{2006A&A...452..459H}
H{\"o}nig S.~F., Beckert T., Ohnaka K., Weigelt G., 2006, A{\&}A, 452, 459

\bibitem[{Hopkins {et~al}\mbox{.}(2008)Hopkins, Hernquist, Cox, \& Kere{\v
  s}}]{2008ApJS..175..356H}
Hopkins P.~F., Hernquist L., Cox T.~J., Kere{\v s} D., 2008, ApJS, 175, 356

\bibitem[{Hopkins \& Quataert(2010)}]{2010MNRAS.407.1529H}
Hopkins P.~F., Quataert E., 2010, MNRAS, 407, 1529

\bibitem[{Hopkins, Richards \& Hernquist(2007)Hopkins, Richards, \&
  Hernquist}]{2007ApJ...654..731H}
Hopkins P.~F., Richards G.~T., Hernquist L., 2007, ApJ, 654, 731

\bibitem[{Keating {et~al}\mbox{.}(2012)Keating, Everett, Gallagher, \&
  Deo}]{2012ApJ...749...32K}
Keating S.~K., Everett J.~E., Gallagher S.~C., Deo R.~P., 2012, ApJ, 749, 32

\bibitem[{King(2010)}]{2010MNRAS.408L..95K}
King A.~R., 2010, MNRAS, 408, L95

\bibitem[{Kinney {et~al}\mbox{.}(2000)Kinney, Schmitt, Clarke, Pringle,
  Ulvestad, \& Antonucci}]{2000ApJ...537..152K}
Kinney A.~L., Schmitt H.~R., Clarke C.~J., Pringle J.~E., Ulvestad J.~S.,
  Antonucci R. R.~J., 2000, ApJ, 537, 152

\bibitem[{Klypin, Trujillo-Gomez \& Primack(2011)Klypin, Trujillo-Gomez, \&
  Primack}]{2011ApJ...740..102K}
Klypin A.~A., Trujillo-Gomez S., Primack J., 2011, ApJ, 740, 102

\bibitem[{Komatsu {et~al}\mbox{.}(2011)Komatsu, Smith, Dunkley, Bennett, Gold,
  Hinshaw, Jarosik, Larson, Nolta, Page, Spergel, Halpern, Hill, Kogut, Limon,
  Meyer, Odegard, Tucker, Weiland, Wollack, \& Wright}]{2011ApJS..192...18K}
Komatsu E. {et~al.}, 2011, ApJS, 192, 18

\bibitem[{Kravtsov \& Klypin(1999)}]{1999ApJ...520..437K}
Kravtsov A.~V., Klypin A.~A., 1999, ApJ, 520, 437

\bibitem[{Krolik \& Begelman(1988)}]{1988ApJ...329..702K}
Krolik J.~H., Begelman M.~C., 1988, ApJ, 329, 702

\bibitem[{Krumpe, Miyaji \& Coil(2010)Krumpe, Miyaji, \&
  Coil}]{2010ApJ...713..558K}
Krumpe M., Miyaji T., Coil A.~L., 2010, ApJ, 713, 558

\bibitem[{La~Franca {et~al}\mbox{.}(2005)La~Franca, Fiore, Comastri, Perola,
  Sacchi, Brusa, Cocchia, Feruglio, Matt, Vignali, Carangelo, Ciliegi,
  Lamastra, Maiolino, Mignoli, Molendi, \& Puccetti}]{2005ApJ...635..864L}
La~Franca F. {et~al.}, 2005, ApJ, 635, 864

\bibitem[{Lacy {et~al}\mbox{.}(2013)Lacy, Ridgway, Gates, Nielsen, Petric,
  Sajina, Urrutia, Cox~Drews, Harrison, Seymour, \&
  Storrie-Lombardi}]{2013ApJS..208...24L}
Lacy M. {et~al.}, 2013, ApJS, 208, 24

\bibitem[{Lacy {et~al}\mbox{.}(2004)Lacy, Storrie-Lombardi, Sajina, Appleton,
  Armus, Chapman, Choi, Fadda, Fang, Frayer, Heinrichsen, Helou, Im, Marleau,
  Masci, Shupe, Soifer, Surace, Teplitz, Wilson, \& Yan}]{2004ApJS..154..166L}
Lacy M. {et~al.}, 2004, ApJS, 154, 166

\bibitem[{Landy \& Szalay(1993)}]{1993ApJ...412...64L}
Landy S.~D., Szalay A.~S., 1993, ApJ, 412, 64

\bibitem[{Lawrence(1991)}]{1991MNRAS.252..586L}
Lawrence A., 1991, MNRAS, 252, 586

\bibitem[{Marinucci {et~al}\mbox{.}(2015)Marinucci, Bianchi, Matt, Alexander,
  Balokovi{\'c}, Bauer, Brandt, Gandhi, Guainazzi, Harrison, Iwasawa, Koss,
  Madsen, Nicastro, Puccetti, Ricci, Stern, \& Walton}]{2015arXiv151103503M}
Marinucci A. {et~al.}, 2015, arXiv, 3503

\bibitem[{Marinucci {et~al}\mbox{.}(2012)Marinucci, Bianchi, Nicastro, Matt, \&
  Goulding}]{2012ApJ...748..130M}
Marinucci A., Bianchi S., Nicastro F., Matt G., Goulding A.~D., 2012, ApJ, 748,
  130

\bibitem[{Mateos {et~al}\mbox{.}(2013)Mateos, Alonso-Herrero, Carrera, Blain,
  Severgnini, Caccianiga, \& Ruiz}]{2013MNRAS.434..941M}
Mateos S., Alonso-Herrero A., Carrera F.~J., Blain A., Severgnini P.,
  Caccianiga A., Ruiz A., 2013, MNRAS, 434, 941

\bibitem[{Mendez {et~al}\mbox{.}(2015)Mendez, Coil, Aird, Skibba,
  Diamond-Stanic, Moustakas, Blanton, Cool, Eisenstein, Wong, \&
  Zhu}]{2015arXiv150406284M}
Mendez A.~J. {et~al.}, 2015, arXiv, 6284

\bibitem[{Merloni {et~al}\mbox{.}(2014)Merloni, Bongiorno, Brusa, Iwasawa,
  Mainieri, Magnelli, Salvato, Berta, Cappelluti, Comastri, Fiore, Gilli,
  Koekemoer, Le~Floc'h, Lusso, Lutz, Miyaji, Pozzi, Riguccini, Rosario,
  Silverman, Symeonidis, Treister, Vignali, \& Zamorani}]{2014MNRAS.437.3550M}
Merloni A. {et~al.}, 2014, MNRAS, 437, 3550

\bibitem[{Miyaji {et~al}\mbox{.}(2011)Miyaji, Krumpe, Coil, \&
  Aceves}]{2011ApJ...726...83M}
Miyaji T., Krumpe M., Coil A.~L., Aceves H., 2011, ApJ, 726, 83

\bibitem[{Moran {et~al}\mbox{.}(2000)Moran, Barth, Kay, \&
  Filippenko}]{2000ApJ...540L..73M}
Moran E.~C., Barth A.~J., Kay L.~E., Filippenko A.~V., 2000, ApJ, 540, L73

\bibitem[{Mullaney {et~al}\mbox{.}(2011)Mullaney, Alexander, Goulding, \&
  Hickox}]{2011MNRAS.414.1082M}
Mullaney J.~R., Alexander D.~M., Goulding A.~D., Hickox R.~C., 2011, MNRAS,
  414, 1082

\bibitem[{Myers {et~al}\mbox{.}(2007)Myers, Brunner, Nichol, Richards,
  Schneider, \& Bahcall}]{2007ApJ...658...85M}
Myers A.~D., Brunner R.~J., Nichol R.~C., Richards G.~T., Schneider D.~P.,
  Bahcall N.~A., 2007, ApJ, 658, 85

\bibitem[{Nenkova {et~al}\mbox{.}(2008{\natexlab{a}})Nenkova, Sirocky,
  Ivezi{\'c}, \& Elitzur}]{2008ApJ...685..147N}
Nenkova M., Sirocky M.~M., Ivezi{\'c} {\v Z}., Elitzur M., 2008{\natexlab{a}},
  ApJ, 685, 147

\bibitem[{Nenkova {et~al}\mbox{.}(2008{\natexlab{b}})Nenkova, Sirocky, Nikutta,
  Ivezi{\'c}, \& Elitzur}]{2008ApJ...685..160N}
Nenkova M., Sirocky M.~M., Nikutta R., Ivezi{\'c} {\v Z}., Elitzur M.,
  2008{\natexlab{b}}, ApJ, 685, 160

\bibitem[{Netzer(2015)}]{2015ARA&A..53..365N}
Netzer H., 2015, ARA{\&}A, 53, 365

\bibitem[{Padmanabhan {et~al}\mbox{.}(2009)Padmanabhan, White, Norberg, \&
  Porciani}]{2009MNRAS.397.1862P}
Padmanabhan N., White M., Norberg P., Porciani C., 2009, MNRAS, 397, 1862

\bibitem[{Planck\hspace{0.1cm}Collaboration
  {et~al}\mbox{.}(2014)Planck\hspace{0.1cm}Collaboration, Ade, Aghanim,
  Armitage-Caplan, Arnaud, Ashdown, Atrio-Barandela, Aumont, Baccigalupi,
  Banday, Barreiro, Bartlett, Basak, Battaner, Benabed, Beno{\^\i}t,
  Benoit-L{\'e}vy, Bernard, Bersanelli, Bielewicz, Bobin, Bock, Bonaldi,
  Bonavera, Bond, Borrill, Bouchet, Bridges, Bucher, Burigana, Butler, Cardoso,
  Catalano, Challinor, Chamballu, Chiang, Chiang, Christensen, Church,
  Clements, Colombi, Colombo, Couchot, Coulais, Crill, Curto, Cuttaia, Danese,
  Davies, Davis, de~Bernardis, de~Rosa, De~Zotti, D{\'e}chelette, Delabrouille,
  Delouis, D{\'e}sert, Dickinson, Diego, Dole, Donzelli, Dor{\'e}, Douspis,
  Dunkley, Dupac, Efstathiou, En{\ss}lin, Eriksen, Finelli, Forni, Frailis,
  Franceschi, Galeotta, Ganga, Giard, Giardino, Giraud-H{\'e}raud,
  Gonz{\'a}lez-Nuevo, G{\'o}rski, Gratton, Gregorio, Gruppuso, Gudmundsson,
  Hansen, Hanson, Harrison, Henrot-Versill{\'e}, Hern{\'a}ndez-Monteagudo,
  Herranz, Hildebrandt, Hivon, Ho, Hobson, Holmes, Hornstrup, Hovest,
  Huffenberger, Jaffe, Jaffe, Jones, Juvela, Keih{\"a}nen, Keskitalo, Kisner,
  Kneissl, Knoche, Knox, Kunz, Kurki-Suonio, Lagache, L{\"a}hteenm{\"a}ki,
  Lamarre, Lasenby, Laureijs, Lavabre, Lawrence, Leahy, Leonardi,
  Le{\'o}n-Tavares, Lesgourgues, Lewis, Liguori, Lilje, Linden-V{\o}rnle,
  L{\'o}pez-Caniego, Lubin, Mac{\'\i}as-P{\'e}rez, Maffei, Maino, Mandolesi,
  Mangilli, Maris, Marshall, Martin, Mart{\'\i}nez-Gonz{\'a}lez, Masi,
  Massardi, Matarrese, Matthai, Mazzotta, Melchiorri, Mendes, Mennella,
  Migliaccio, Mitra, Miville-Desch{\^e}nes, Moneti, Montier, Morgante,
  Mortlock, Moss, Munshi, Murphy, Naselsky, Nati, Natoli, Netterfield,
  N{\o}rgaard-Nielsen, Noviello, Novikov, Novikov, Osborne, Oxborrow, Paci,
  Pagano, Pajot, Paoletti, Partridge, Pasian, Patanchon, Perdereau, Perotto,
  Perrotta, Piacentini, Piat, Pierpaoli, Pietrobon, Plaszczynski,
  Pointecouteau, Polenta, Ponthieu, Popa, Poutanen, Pratt, Pr{\'e}zeau, Prunet,
  Puget, Pullen, Rachen, Rebolo, Reinecke, Remazeilles, Renault, Ricciardi,
  Riller, Ristorcelli, Rocha, Rosset, Roudier, Rowan-Robinson,
  Rubi{\~n}o-Mart{\'\i}n, Rusholme, Sandri, Santos, Savini, Scott, Seiffert,
  Shellard, Smith, Spencer, Starck, Stolyarov, Stompor, Sudiwala, Sunyaev,
  Sureau, Sutton, Suur-Uski, Sygnet, Tauber, Tavagnacco, Terenzi, Toffolatti,
  Tomasi, Tristram, Tucci, Tuovinen, Umana, Valenziano, Valiviita, Van~Tent,
  Vielva, Villa, Vittorio, Wade, Wandelt, White, White, Yvon, Zacchei, \&
  Zonca}]{2014A&A...571A..17P}
Planck\hspace{0.1cm}Collaboration {et~al.}, 2014, A{\&}A, 571, A17

\bibitem[{Planck\hspace{0.1cm}Collaboration
  {et~al}\mbox{.}(2015)Planck\hspace{0.1cm}Collaboration, Ade, Aghanim, Arnaud,
  Ashdown, Aumont, Baccigalupi, Banday, Barreiro, Bartlett, Bartolo, Battaner,
  Benabed, Beno{\^\i}t, Benoit-L{\'e}vy, Bernard, Bersanelli, Bielewicz,
  Bonaldi, Bonavera, Bond, Borrill, Bouchet, Boulanger, Bucher, Burigana,
  Butler, Calabrese, Cardoso, Catalano, Challinor, Chamballu, Chiang,
  Christensen, Church, Clements, Colombi, Colombo, Combet, Couchot, Coulais,
  Crill, Curto, Cuttaia, Danese, Davies, Davis, de~Bernardis, de~Rosa,
  De~Zotti, Delabrouille, D{\'e}sert, Diego, Dole, Donzelli, Dor{\'e}, Douspis,
  Ducout, Dunkley, Dupac, Efstathiou, Elsner, En{\ss}lin, Eriksen, Fergusson,
  Finelli, Forni, Frailis, Fraisse, Franceschi, Frejsel, Galeotta, Galli,
  Ganga, Giard, Giraud-H{\'e}raud, Gjerl{\o}w, Gonz{\'a}lez-Nuevo, G{\'o}rski,
  Gratton, Gregorio, Gruppuso, Gudmundsson, Hansen, Hanson, Harrison,
  Henrot-Versill{\'e}, Hern{\'a}ndez-Monteagudo, Herranz, Hildebrandt, Hivon,
  Hobson, Holmes, Hornstrup, Hovest, Huffenberger, Hurier, Jaffe, Jaffe, Jones,
  Juvela, Keih{\"a}nen, Keskitalo, Kisner, Kneissl, Knoche, Kunz, Kurki-Suonio,
  Lagache, L{\"a}hteenm{\"a}ki, Lamarre, Lasenby, Lattanzi, Lawrence, Leonardi,
  Lesgourgues, Levrier, Lewis, Liguori, Lilje, Linden-V{\o}rnle,
  L{\'o}pez-Caniego, Lubin, Mac{\'\i}as-P{\'e}rez, Maggio, Maino, Mandolesi,
  Mangilli, Martin, Mart{\'\i}nez-Gonz{\'a}lez, Masi, Matarrese, Mazzotta,
  McGehee, Meinhold, Melchiorri, Mendes, Mennella, Migliaccio, Mitra,
  Miville-Desch{\^e}nes, Moneti, Montier, Morgante, Mortlock, Moss, Munshi,
  Murphy, Naselsky, Nati, Natoli, Netterfield, N{\o}rgaard-Nielsen, Noviello,
  Novikov, Novikov, Oxborrow, Paci, Pagano, Pajot, Paoletti, Pasian, Patanchon,
  Perdereau, Perotto, Perrotta, Pettorino, Piacentini, Piat, Pierpaoli,
  Pietrobon, Plaszczynski, Pointecouteau, Polenta, Popa, Pratt, Pr{\'e}zeau,
  Prunet, Puget, Rachen, Reach, Rebolo, Reinecke, Remazeilles, Renault, Renzi,
  Ristorcelli, Rocha, Rosset, Rossetti, Roudier, Rowan-Robinson,
  Rubi{\~n}o-Mart{\'\i}n, Rusholme, Sandri, Santos, Savelainen, Savini, Scott,
  Seiffert, Shellard, Spencer, Stolyarov, Stompor, Sudiwala, Sunyaev, Sutton,
  Suur-Uski, Sygnet, Tauber, Terenzi, Toffolatti, Tomasi, Tristram, Tucci,
  Tuovinen, Valenziano, Valiviita, Van~Tent, Vielva, Villa, Wade, Wandelt,
  Wehus, White, Yvon, Zacchei, \& Zonca}]{PlanckCollaboration:2015tp}
Planck\hspace{0.1cm}Collaboration {et~al.}, 2015, arXiv

\bibitem[{Polletta {et~al}\mbox{.}(2008)Polletta, Weedman, H{\"o}nig, Lonsdale,
  Smith, \& Houck}]{2008ApJ...675..960P}
Polletta M., Weedman D., H{\"o}nig S., Lonsdale C.~J., Smith H.~E., Houck J.,
  2008, ApJ, 675, 960

\bibitem[{Porciani, Magliocchetti \& Norberg(2004)Porciani, Magliocchetti, \&
  Norberg}]{2004MNRAS.355.1010P}
Porciani C., Magliocchetti M., Norberg P., 2004, MNRAS, 355, 1010

\bibitem[{Prada {et~al}\mbox{.}(2012)Prada, Klypin, Cuesta, Betancort-Rijo, \&
  Primack}]{2012MNRAS.423.3018P}
Prada F., Klypin A.~A., Cuesta A.~J., Betancort-Rijo J.~E., Primack J., 2012,
  MNRAS, 423, 3018

\bibitem[{Richardson {et~al}\mbox{.}(2013)Richardson, Chatterjee, Zheng, Myers,
  \& Hickox}]{2013ApJ...774..143R}
Richardson J., Chatterjee S., Zheng Z., Myers A.~D., Hickox R., 2013, ApJ, 774,
  143

\bibitem[{Richardson {et~al}\mbox{.}(2012)Richardson, Zheng, Chatterjee, Nagai,
  \& Shen}]{2012ApJ...755...30R}
Richardson J., Zheng Z., Chatterjee S., Nagai D., Shen Y., 2012, ApJ, 755, 30

\bibitem[{Ross {et~al}\mbox{.}(2009)Ross, Shen, Strauss, Vanden~Berk, Connolly,
  Richards, Schneider, Weinberg, Hall, Bahcall, \&
  Brunner}]{2009ApJ...697.1634R}
Ross N.~P. {et~al.}, 2009, ApJ, 697, 1634

\bibitem[{Sanders {et~al}\mbox{.}(1988)Sanders, Soifer, Elias, Madore,
  Matthews, Neugebauer, \& Scoville}]{1988ApJ...325...74S}
Sanders D.~B., Soifer B.~T., Elias J.~H., Madore B.~F., Matthews K., Neugebauer
  G., Scoville N.~Z., 1988, ApJ, 325, 74

\bibitem[{Shankar {et~al}\mbox{.}(2006)Shankar, Shankar, Lapi, Lapi, Salucci,
  Salucci, De~Zotti, De~Zotti, Danese, \& Danese}]{2006ApJ...643...14S}
Shankar F. {et~al.}, 2006, ApJ, 643, 14

\bibitem[{Shen {et~al}\mbox{.}(2013)Shen, McBride, White, Zheng, Myers, Guo,
  Kirkpatrick, Padmanabhan, Parejko, Ross, Schlegel, Schneider, Streblyanska,
  Swanson, Zehavi, Pan, Bizyaev, Brewington, Ebelke, Malanushenko,
  Malanushenko, Oravetz, Simmons, \& Snedden}]{2013ApJ...778...98S}
Shen Y. {et~al.}, 2013, ApJ, 778, 98

\bibitem[{Sherwin {et~al}\mbox{.}(2012)Sherwin, Das, Hajian, Addison, Bond,
  Crichton, Devlin, Dunkley, Gralla, Halpern, Hill, Hincks, Hughes,
  Huffenberger, Hlozek, Kosowsky, Louis, Marriage, Marsden, Menanteau, Moodley,
  Niemack, Page, Reese, Sehgal, Sievers, Sif{\'o}n, Spergel, Staggs, Switzer,
  \& Wollack}]{2012PhRvD..86h3006S}
Sherwin B.~D. {et~al.}, 2012, PhRvD, 86, 83006

\bibitem[{Sheth, Mo \& Tormen(2001)Sheth, Mo, \& Tormen}]{2001MNRAS.323....1S}
Sheth R.~K., Mo H.~J., Tormen G., 2001, MNRAS, 323, 1

\bibitem[{Sijacki {et~al}\mbox{.}(2015)Sijacki, Vogelsberger, Genel, Springel,
  Torrey, Snyder, Nelson, \& Hernquist}]{2015MNRAS.452..575S}
Sijacki D., Vogelsberger M., Genel S., Springel V., Torrey P., Snyder G.~F.,
  Nelson D., Hernquist L., 2015, MNRAS, 452, 575

\bibitem[{Springel {et~al}\mbox{.}(2005)Springel, White, Jenkins, Frenk,
  Yoshida, Gao, Navarro, Thacker, Croton, Helly, Peacock, Cole, Thomas,
  Couchman, Evrard, Colberg, \& Pearce}]{2005Natur.435..629S}
Springel V. {et~al.}, 2005, Nature, 435, 629

\bibitem[{Stalevski {et~al}\mbox{.}(2012)Stalevski, Fritz, Baes, Nakos, \&
  Popovic}]{2012MNRAS.420.2756S}
Stalevski M., Fritz J., Baes M., Nakos T., Popovic L.~C., 2012, MNRAS, 420,
  2756

\bibitem[{Starikova {et~al}\mbox{.}(2011)Starikova, Cool, Eisenstein, Forman,
  Jones, Hickox, Kenter, Kochanek, Kravtsov, Murray, \&
  Vikhlinin}]{2011ApJ...741...15S}
Starikova S. {et~al.}, 2011, ApJ, 741, 15

\bibitem[{Stern {et~al}\mbox{.}(2012)Stern, Assef, Benford, Blain, Cutri, Dey,
  Eisenhardt, Griffith, Jarrett, Lake, Masci, Petty, Stanford, Tsai, Wright,
  Yan, Harrison, \& Madsen}]{2012ApJ...753...30S}
Stern D. {et~al.}, 2012, ApJ, 753, 30

\bibitem[{Stern {et~al}\mbox{.}(2005)Stern, Eisenhardt, Gorjian, Kochanek,
  Caldwell, Eisenstein, Brodwin, Brown, Cool, Dey, Green, Jannuzi, Murray,
  Pahre, \& Willner}]{Stern:2005p2563}
Stern D. {et~al.}, 2005, ApJ, 631, 163

\bibitem[{Swanson {et~al}\mbox{.}(2008)Swanson, Tegmark, Hamilton, \&
  Hill}]{2008MNRAS.387.1391S}
Swanson M. E.~C., Tegmark M., Hamilton A. J.~S., Hill J.~C., 2008, MNRAS, 387,
  1391

\bibitem[{Tinker {et~al}\mbox{.}(2010)Tinker, Robertson, Kravtsov, Klypin,
  Warren, Yepes, \& Gottlober}]{2010ApJ...724..878T}
Tinker J.~L., Robertson B.~E., Kravtsov A.~V., Klypin A., Warren M.~S., Yepes
  G., Gottlober S., 2010, ApJ, 724, 878

\bibitem[{Tinker {et~al}\mbox{.}(2005)Tinker, Weinberg, Zheng, \&
  Zehavi}]{2005ApJ...631...41T}
Tinker J.~L., Weinberg D.~H., Zheng Z., Zehavi I., 2005, ApJ, 631, 41

\bibitem[{Treister, Krolik \& Dullemond(2008)Treister, Krolik, \&
  Dullemond}]{2008ApJ...679..140T}
Treister E., Krolik J.~H., Dullemond C., 2008, ApJ, 679, 140

\bibitem[{Ueda {et~al}\mbox{.}(2003)Ueda, Akiyama, Ohta, \&
  Miyaji}]{2003ApJ...598..886U}
Ueda Y., Akiyama M., Ohta K., Miyaji T., 2003, ApJ, 598, 886

\bibitem[{Urry \& Padovani(1995)}]{Urry:1995p507}
Urry C.~M., Padovani P., 1995, Publications of the Astronomical Society of the
  Pacific, 107, 803

\bibitem[{Vale \& Ostriker(2004)}]{2004MNRAS.353..189V}
Vale A., Ostriker J.~P., 2004, MNRAS, 353, 189

\bibitem[{van Engelen {et~al}\mbox{.}(2012)van Engelen, Keisler, Zahn, Aird,
  Benson, Bleem, Carlstrom, Chang, Cho, Crawford, Crites, de~Haan, Dobbs,
  Dudley, George, Halverson, Holder, Holzapfel, Hoover, Hou, Hrubes, Joy, Knox,
  Lee, Leitch, Lueker, Luong-Van, McMahon, Mehl, Meyer, Millea, Mohr, Montroy,
  Natoli, Padin, Plagge, Pryke, Reichardt, Ruhl, Sayre, Schaffer, Shaw,
  Shirokoff, Spieler, Staniszewski, Stark, Story, Vanderlinde, Vieira, \&
  Williamson}]{2012ApJ...756..142V}
van Engelen A. {et~al.}, 2012, ApJ, 756, 142

\bibitem[{Vogelsberger {et~al}\mbox{.}(2014)Vogelsberger, Genel, Springel,
  Torrey, Sijacki, Xu, Snyder, Nelson, \& Hernquist}]{2014MNRAS.444.1518V}
Vogelsberger M. {et~al.}, 2014, MNRAS, 444, 1518

\bibitem[{Wang \& Brunner(2014)}]{2014MNRAS.444.2854W}
Wang Y., Brunner R.~J., 2014, MNRAS, 444, 2854

\bibitem[{Werner {et~al}\mbox{.}(2004)Werner, Roellig, Low, Rieke, Rieke,
  Hoffmann, Young, Houck, Brandl, Fazio, Hora, Gehrz, Helou, Soifer, Stauffer,
  Keene, Eisenhardt, Gallagher, Gautier, Irace, Lawrence, Simmons, Van~Cleve,
  Jura, Wright, \& Cruikshank}]{2004ApJS..154....1W}
Werner M.~W. {et~al.}, 2004, The Astrophysical Journal Supplement Series, 154,
  1

\bibitem[{White {et~al}\mbox{.}(2012)White, Myers, Ross, Schlegel, Hennawi,
  Shen, McGreer, Strauss, Bolton, Bovy, Fan, Miralda-Escude,
  Palanque-Delabrouille, Paris, Petitjean, Schneider, Viel, Weinberg, Yeche,
  Zehavi, Pan, Snedden, Bizyaev, Brewington, Brinkmann, Malanushenko,
  Malanushenko, Oravetz, Simmons, Sheldon, \& Weaver}]{2012MNRAS.424..933W}
White M. {et~al.}, 2012, MNRAS, 424, 933

\bibitem[{Wright {et~al}\mbox{.}(2010)Wright, Eisenhardt, Mainzer, Ressler,
  Cutri, Jarrett, Kirkpatrick, Padgett, McMillan, Skrutskie, Stanford, Cohen,
  Walker, Mather, Leisawitz, Gautier, McLean, Benford, Lonsdale, Blain, Mendez,
  Irace, Duval, Liu, Royer, Heinrichsen, Howard, Shannon, Kendall, Walsh,
  Larsen, Cardon, Schick, Schwalm, Abid, Fabinsky, Naes, \&
  Tsai}]{2010AJ....140.1868W}
Wright E.~L. {et~al.}, 2010, AJ, 140, 1868

\bibitem[{York {et~al}\mbox{.}(2000)York, Adelman, Anderson, Anderson, Annis,
  Bahcall, Bakken, Barkhouser, Bastian, Berman, Boroski, Bracker, Briegel,
  Briggs, Brinkmann, Brunner, Burles, Carey, Carr, Castander, Chen, Colestock,
  Connolly, Crocker, Csabai, Czarapata, Davis, Doi, Dombeck, Eisenstein,
  Ellman, Elms, Evans, Fan, Federwitz, Fiscelli, Friedman, Frieman, Fukugita,
  Gillespie, Gunn, Gurbani, de~Haas, Haldeman, Harris, Hayes, Heckman,
  Hennessy, Hindsley, Holm, Holmgren, Huang, Hull, Husby, Ichikawa, Ichikawa,
  Ivezi{\'c}, Kent, Kim, Kinney, Klaene, Kleinman, Kleinman, Knapp, Korienek,
  Kron, Kunszt, Lamb, Lee, Leger, Limmongkol, Lindenmeyer, Long, Loomis,
  Loveday, Lucinio, Lupton, MacKinnon, Mannery, Mantsch, Margon, McGehee,
  McKay, Meiksin, Merelli, Monet, Munn, Narayanan, Nash, Neilsen, Neswold,
  Newberg, Nichol, Nicinski, Nonino, Okada, Okamura, Ostriker, Owen, Pauls,
  Peoples, Peterson, Petravick, Pier, Pope, Pordes, Prosapio, Rechenmacher,
  Quinn, Richards, Richmond, Rivetta, Rockosi, Ruthmansdorfer, Sandford,
  Schlegel, Schneider, Sekiguchi, Sergey, Shimasaku, Siegmund, Smee, Smith,
  Snedden, Stone, Stoughton, Strauss, Stubbs, SubbaRao, Szalay, Szapudi,
  Szokoly, Thakar, Tremonti, Tucker, Uomoto, Vanden~Berk, Vogeley, Waddell,
  Wang, Watanabe, Weinberg, Yanny, Yasuda, \&
  collaboration}]{2000AJ....120.1579Y}
York D.~G. {et~al.}, 2000, AJ, 120, 1579

\bibitem[{Zehavi {et~al}\mbox{.}(2011)Zehavi, Zheng, Weinberg, Blanton,
  Bahcall, Berlind, Brinkmann, Frieman, Gunn, Lupton, Nichol, Percival,
  Schneider, Skibba, Strauss, Tegmark, \& York}]{2011ApJ...736...59Z}
Zehavi I. {et~al.}, 2011, ApJ, 736, 59

\bibitem[{Zhang {et~al}\mbox{.}(2009)Zhang, Soria, Zhang, Swartz, \&
  Liu}]{2009ApJ...699..281Z}
Zhang W.~M., Soria R., Zhang S.-N., Swartz D.~A., Liu J.~F., 2009, ApJ, 699,
  281

\bibitem[{Zheng {et~al}\mbox{.}(2005)Zheng, Berlind, Weinberg, Benson, Baugh,
  Cole, Dav{\'e}, Frenk, Katz, \& Lacey}]{2005ApJ...633..791Z}
Zheng Z. {et~al.}, 2005, ApJ, 633, 791

\end{thebibliography}

\label{lastpage}

\end{document}